\documentstyle[prd,aps,floats]{revtex}

\begin{document}
\preprint{SUSSEX-AST 98/?-?, gr-qc/9811040} \draft \tighten
%
%
\input epsf
\renewcommand{\topfraction}{0.8} \renewcommand{\bottomfraction}{0.8}

\title{Early cosmology and the stochastic gravitational wave
  background} \author{Lu\'{\i}s E.~Mendes and Andrew R.~Liddle}
\address{Astronomy Centre, University of Sussex, Falmer, Brighton BN1
  9QJ,~~U.~K.\\and\\Astrophysics Group, The Blackett Laboratory,
  Imperial College, London SW7 2BZ,~~U.~K.~~~(present address)}
\date{\today} \maketitle
\begin{abstract}
  The epoch when the Universe had a temperature higher than a ${\rm
    GeV}$ is long before any time at which we have reliable
  observations constraining the cosmological evolution.  For example,
  the occurrence of a second burst of inflation (sometimes called
  thermal inflation) at a lower energy scale than standard inflation,
  or a short epoch of early matter domination, cannot be ruled out by
  present cosmological data.  The cosmological stochastic
  gravitational wave background, on scales accessible to
  interferometer detection, is sensitive to non-standard cosmologies
  of this type.  We consider the implications of such alternative
  models both for ground-based experiments such as {\sc ligo} and
  space-based proposals such as {\sc lisa}. We show that a
  second burst of inflation leads to a scale-dependent reduction in the
  spectrum. Applied to conventional inflation, this further reduces an already
  disappointingly low signal. In the pre big bang scenario, where a
  much more potent signal is possible, the amplitude is reduced
  but the background remains observable by {\sc lisa} in certain
  parameter space regions.  In each case, a second epoch of inflation
  induces oscillatory features into the spectrum in a manner analogous
  to the acoustic peaks in the density perturbation spectrum.  On {\sc
    ligo} scales, perturbations can only survive through thermal
  inflation with detectable amplitudes if their amplitudes were at one
  time so large that linear perturbation theory is inadequate.
  Although for an epoch of early matter domination the reduction in
  the expected signal is not as large as the one caused by a second
  burst of inflation, the detection in the context of the pre big bang 
  scenario may not be possible since the spectrum peaks around the
  {\sc ligo} frequency window and for lower frequencies behaves as $f^{3}$.
\end{abstract}

\pacs{PACS numbers: 04.30.-w, 98.80.Cq \hspace*{5.6cm} Preprint
  gr-qc/9811040}

\vskip2pc

\section{Introduction}

A stochastic gravitational wave background is potentially a powerful
source of cosmological information \cite{revs}, since gravitational
waves decoupled from the matter in the Universe at an extremely early
stage.  Various ways exist in which such a background might be
generated, including cosmological inflation, collision of bubbles
during a phase transition, and motions of topological defects.  We
will be concentrating on the first of these.

There are currently four large-scale ground-based interferometers
under construction, namely {\sc ligo}, {\sc virgo}, {\sc geo} and {\sc
  tama}, and a proposal for a space-based detector {\sc lisa} is also
under consideration; see Ref.~\cite{revs} for reviews of these
experiments.  Unlike measures of density perturbations, these
experiments probe frequencies which are high enough that the
gravitational waves have had a wavelength less than the Hubble radius
since very early in the history of the Universe.  Assuming the
standard cosmology, where the Universe was radiation dominated from an
early stage, giving way to matter domination relatively recently, the
time when a comoving frequency $f_*$ equalled the Hubble scale is
given by
\begin{equation}
\label{fs}
\frac{f_*}{f_0} = \frac{H_* a_*}{H_0 a_0} 
        \approx \frac{T_*}{T_{{\rm eq}}} \, z_{{\rm eq}}^{1/2} \,,
\end{equation}
where $f_0 = a_0 H_0 = 3h \times 10^{-18}$ Hz is the mode that is just
re-entering the Hubble radius today, $T$ is the temperature and
$z_{{\rm eq}}=24\,000\,\Omega_0 h^2$ the redshift of matter--radiation
equality.  Here $\Omega_0$ and $h$ are the density parameter and
Hubble constant in the usual units. Comoving units have been
normalized to match present physical units.  Since $T_{\rm eq} =
24\,000\,\Omega_0 h^2 \, T_0 \approx 1{\rm eV}$, we have
\begin{equation}
\label{fs2}
{f_* \over 100\, {\rm Hz}} \approx {T_* \over 10^9\, {\rm GeV}} \,,
\end{equation} 
where $100\, {\rm Hz}$ is the typical frequency band probed by
ground-based interferometers.

The epoch when the Universe had a temperature higher than a GeV is
long before any time at which we have reliable observations
constraining the cosmological evolution.  Because interesting
wavelengths have been less than the Hubble radius since then, the
amplitude of the stochastic background will be sensitive to whatever
the actual cosmological evolution is.  A different cosmological model
will have two principal effects: first, it will modify the
relationship between scales given by Eq.~(\ref{fs2}) for the standard
cosmology, and secondly it can alter the amplitude.  Only those scales
which are larger than the Hubble radius will have their amplitude
unchanged in alternative cosmologies.

To highlight this effect, we shall consider two alternative scenarios
to the standard cosmology.  In the first we assume a second short
burst of inflation at a lower energy scale than standard inflation,
often called thermal inflation~\cite{LS}.  Such a short stage of
inflation may be desirable to get rid of unwanted relic particles
which could otherwise have dramatic consequences on the subsequent
evolution of the Universe~\cite{LS}.  This second burst of inflation
has in general no connection to the usual inflationary epoch which is
assumed responsible for the generation of the density perturbations
and gravitational waves leading to structure formation and microwave
anisotropies.  It is normally imagined to begin at a temperature
around $10^7 \, {\rm GeV}$, and continue until about $10^3 \, {\rm
  GeV}$ (the supersymmetry scale), giving a total of $\ln 10^4 \simeq
10$ $e$-foldings of inflation.  If this second burst of inflation
occurs after the scales of interest (i.e.  the scales accessible to
{\sc ligo} and {\sc lisa}) are within the horizon, then the stochastic
background is altered.

The second alternative is the recently discussed possibility that the
Universe becomes temporarily dominated by some long-lived massive
particle (for example the moduli fields of string theory), inducing an
early period of matter domination which ends when the particle finally
decays to restore the usual radiation-dominated Universe (see
Ref.~\cite{early-matter} and references therein).  Another possible
origin for a short period of early matter domination lies with the QCD
phase transition, as was recently discussed by Schwarz~\cite{Schwarz}.

To make the discussion concrete, we require a model for the initial
gravitational wave spectrum.  We consider two somewhat different
models for the actual origin of the stochastic background.  The first
is conventional inflation, which is the standard case but which gives
a disappointingly low signal even in the standard cosmology.  The
second is an unusual variant on inflation known as the pre big bang
(PBB) scenario~\cite{pbb1,pbb2-frame-equiv}.  This is a much more
speculative idea, but potentially of greater interest as it predicts a
spectrum which rises sharply towards small scales
\cite{pbbgw,mode-eq}, the opposite of the usual inflationary
behaviour, thus creating a realistic prospect of direct detection of
the gravitational wave background \cite{pbbgwdet}.

\section{Cosmology with thermal inflation}

The first modification to the standard Big Bang model we consider is a
second epoch of inflation occurring after the beginning of the
radiation epoch. We assume that the original epoch of inflation
(either conventional or pre big bang) responsible for generating the
gravitational wave background ends at conformal time $\tau_{{\rm r}1}$
(from now on we will always use conformal time), at which point the
radiation era begins. At time $\tau_{{\rm i}2}$ a second era of
inflation starts which ends at $\tau_{{\rm r}2}$; we will model this
era as exponential inflation.  After $\tau_{{\rm r}2}$ the Universe
again becomes radiation dominated and the evolution of the Universe
continues according the standard model with an era of matter
domination starting at $\tau_{\rm m}$ which lasts up to the present.
The scale factor after the original period of inflation is given by
\begin{equation}
  \label{eq:sf_inf}
  a(\tau) = \left\{ 
    \begin{array}{lr}
      {\displaystyle a_2 \left( \tau_2 + \tau \right)}\, ; &  
      \tau_{{\rm r}1} < \tau < \tau_{{\rm i}2} \\[4mm]
      {\displaystyle a_3 \left( \tau_3 - \tau \right)^{-1}} \, ; &   
      \tau_{{\rm i}2} < \tau < \tau_{{\rm r}2} \\[4mm]
      {\displaystyle a_4 \tau} \, ; 
      &  \tau_{{\rm r}2} < \tau < \tau_{{\rm eq}} \\[4mm]
      {\displaystyle a_5 \left(\tau_5 + \tau \right)^2} \, ; &  
      \tau > \tau_{{\rm eq}}
    \end{array}
\right.
\end{equation}
where the constants are determined by requiring the continuity of $a$
and $da/d\tau$ at the transition between two epochs. For the moment we
only want to make an approximate estimate of the modifications
introduced in the spectrum of the stochastic gravitational wave
background and we will not worry about the exact value of these
constants. We parametrize the thermal inflation by the energy scale at
which it starts, $\rho_{{\rm i}2}$, and by the total expansion it
gives rise to, $R_{\rm i}=a_{{\rm r}2}/a_{{\rm i}2}$.

In this model we assume that the first epoch of radiation domination
happens at a temperature of order $10^{16}$ GeV, while the short
period of thermal inflation begins around $10^8$ GeV. We also assume
that during thermal inflation the Universe expands by a factor of
$10^4$ (approximately $10$ $e$-foldings of inflation). This is the
minimal amount of inflation which solves the moduli problem~\cite{LS}.

Normalizing the scale factor at the present as $a_0=1$, we can write
all the relevant parameters in our model in terms of $H_{{\rm i}}$,
$H_{{\rm i}2}$ (the Hubble parameters at the end of the first period
of inflation and at beginning of the second stage of inflation),
$R_{\rm i}$, the present Hubble parameter $H_0$ and the redshift at
the matter--radiation equality $z_{{\rm eq}}$. In terms of these
parameters we can write the Hubble parameter and time at the
matter--radiation equality
\begin{equation}
  \label{eq:eq}
  H_{{\rm eq}} = (1+z_{{\rm eq}})^{3/2} H_0 \,\, , \quad \tau_{{\rm eq}} = 
  \left( 1 + z_{{\rm eq}} \right)^{-1/2} H_0^{-1} \, ,
\end{equation}
the Hubble parameter, redshift and time at the end of the second epoch
of inflation
\begin{equation}
  \label{eq:r2}
  H_{{\rm r}2} = H_{{\rm i}2} \,\, , \quad 
  1+z_{{\rm r}2} = \left( 1+z_{\rm eq}\right)^{1/4} 
  \left( \frac{H_{{\rm i}2}}{H_0} \right)^{1/2} \,\, , \quad
  \tau_{{\rm r}2} = \frac{\left( 1 + z_{{\rm eq}}
      \right)^{1/4}}{(H_{{\rm i}2} H_0)^{1/2}} \, ,
\end{equation}
as well as the redshift and time at the beginning of the second period
of inflation
\begin{equation}
  \label{eq:i2}
   1+z_{{\rm i}2}= \left( 1+z_{\rm eq}\right)^{1/4} R_{\rm i} \left( 
     \frac{H_{{\rm i}2}}{H_0} \right)^{1/2} \,\, , \quad
  \tau_{{\rm i}2} \approx  - \frac{R_{\rm i} \left( 1
    + z_{{\rm eq}} \right)^{1/4}}{(H_{{\rm i}2} H_0)^{1/2}}\, ,
\end{equation}
and at the end of the first epoch of inflation
\begin{equation}
  \label{eq:r1}
   1+z_{{\rm r}1} = \left( 1+z_{\rm eq}\right)^{1/4} R_{\rm i}
  \left( \frac{H_{{\rm r}1}}{H_0} \right)^{1/2}\,\, , \quad \tau_{{\rm
      r}1} \approx  -\frac{2 R_{\rm i} 
    \left( 1 + z_{{\rm eq}} \right)^{1/4}}{(H_{{\rm i}2} H_0)^{1/2}}\, .
\end{equation}

We now describe the physical effects through which thermal inflation
modifies the spectrum. The technical details of the calculation are
quite complex, and we describe them fully in the Appendix; here in the
main text we will restrict ourselves to describing the relevant
physics and exhibiting the results.

On short and long scales things are relatively simple. Modes always
above the horizon cannot change their amplitude, but there is an
apparent effect caused by the change in the correspondence between
comoving scales (i.e.~physical scales during inflation) and present
physical scales; the extra expansion during thermal inflation shifts
the spectrum to long wavelengths. For a flat spectrum this has no
effect, but if the spectrum is tilted this alters the amplitude on a
given present physical scale. This effect may have observational
consequences in the contribution of tensor perturbations to the
anisotropy of the cosmic microwave background radiation as we will
discuss in the following subsection.

A related effect is a redshift in the high-frequency cutoff of the
spectrum corresponding to the horizon size at the end of the main
period of inflation. From Eq.~(\ref{eq:hfc}) we obtain
\begin{equation}
  \label{eq:hfc1}
  f_{{\rm r}1} = \frac{\left( H_{{\rm r}1} H_{0} \right)^{1/2}}{(1+z_{\rm
        eq})^{1/4} R_{\rm i}} \, .
\end{equation}

Very short scale modes, which are always inside the horizon when the
alternate behaviour is occurring, simply suffer an additional
redshift. Modes inside the horizon behave as radiation, with the
energy density decreasing as $a^{-4}$. If thermal inflation occurs it
means there is more total expansion between the initial generation of
the modes and the present, and so their amplitude is suppressed. For
example, with perfect exponential thermal inflation, this suppression
will be $R_{\rm i}^{-4} \sim 10^{-16}$. Had we assumed an epoch of
power-law thermal inflation with $a\propto\tau^{-q}$ ($q \geq 1$),
$f_{{\rm r}1}$ would be redshifted by $R_{\rm i}^{(1+q)/2q}$.
Recalling that during an epoch of thermal inflation the total energy density
redshifts as $a^{2(1/p-1)}$, the suppression in $\Omega_{\rm gw}$
in this case would be by a factor $\rho_{\rm gw}/\rho = R_{\rm
  i}^{-2(1+1/q)}$.

The most interesting regime is the intermediate one.  For waves with
frequencies in the interval $f_{{\rm i}2} < f < f_{{\rm r}2}$, a
curious phenomenon happens: these waves enter the Hubble radius during
the first epoch of radiation but are again pushed outside the horizon
during the subsequent epoch of inflation. While they are inside the
horizon their behaviour is oscillatory.  Hence, when they leave the
horizon, their phases will have a strong dependence on frequency and
this will leave an oscillatory imprint in the spectrum. This is
strongly reminiscent of the acoustic peak structure in the density
perturbation spectrum which leads to oscillations in the radiation
power spectrum, and is arising similarly from the dominance of growing
mode perturbations at horizon entry. Similar oscillations have also
been found in the primordial density perturbation spectrum in the case
of double inflation~\cite{star}.

To make our discussion concrete, we now exhibit results for two
particular models for the initial spectrum.

\subsection{Conventional inflation}

We model conventional inflation as a period of power-law
inflation~\cite{plinf}, corresponding to an initial epoch where the
scale factor behaves as
\begin{equation}
\label{eq:ci}
a(\tau) = a_1 (\tau_1 - \tau)^{-p} \,; \quad -\infty<\tau<\tau_{{\rm r}1} \,.
\end{equation}
where $p \geq 1$ (equality holds for exponential inflation). When we
write the scale factor in comoving time we get $a\propto t^\alpha$
with $p$ and $\alpha$ related by $p=\alpha/(\alpha-1)$.

In this standard inflationary scenario the spectral energy density
parameter for the gravitational waves can be calculated in the
long-wavelength approximation where we assume that $k\tau\ll 1$ (see
the Appendix for the technical details). After a straightforward but
tedious calculation we obtain
\begin{equation}
  \label{eq:bogapprox_expci}
   \Omega_{{\rm gw}} = \left\{
  \begin{array}{lrcl}
    {\displaystyle \frac{2^{3p-2}}{3^p} \frac{\pi^{1-p} \, \Gamma}{p^{2 \left(
          1+p \right)}} \frac{\left( \rho_{\rm c} \, \rho_{\rm
          r}\right)^{\left(1+p\right)/2}}{\rho_{\rm Pl}^p} \left(
      \frac{f_{\rm Pl}}{f}\right)^{2 p}} \quad ; &  
    H_0 & <f< & {\displaystyle H_0 (1+z_{\rm eq})^{1/2}} \\[5mm]
    {\displaystyle \frac{2^{3 p + 1}}{3^{p+1}} \frac{\pi^{2 \left( 1 - 
          p/2 \right)} \, \Gamma}{p^{2 \left( 1 + p \right)}} 
    \frac{\left( \rho_{\rm c}^{p-1} \rho_{\rm r}^{p+1}
      \right)^{1/2}}{\rho_{\rm Pl}^p \left( 1 + z_{\rm eq} 
      \right)^{\left( 1+p\right)/2}} \left(
      \frac{f_{\rm Pl}}{f}\right)^{2 \left( p - 1 \right)}}; 
  \quad & {\displaystyle H_0 
      (1+z_{\rm eq})^{1/2}} & < f < &
      {\displaystyle \frac{\left( H_{\rm i} H_{0} \right)^{1/2}}{(1+z_{\rm
        eq})^{1/4}}} \, .
  \end{array}
  \right.
\end{equation}
where $\Gamma$ is given by
\begin{displaymath}
  \Gamma = \frac{\left[\Gamma\!\left( - \frac{1}{2} - p \right)+2
    \Gamma\!\left(\frac{1}{2} - p
    \right)\right]^2}{\left[ \Gamma\!\left(-\frac{1}{2} - p \right)
    \Gamma\!\left(\frac{1}{2} - p\right)\right]^2} \sec^2\!\left(p \pi\right)
\end{displaymath}
When $p=1$ this gives $\Gamma = 1/\pi$ and we recover the usual result
for exponential inflation.

From the observational point of view, the extra redshift suffered by
waves inside the horizon during the second stage of inflation will
have a devastating effect on the prospects for observing them.
Assuming for instance that during the second epoch of inflation we
have exponential inflation, we will have an extra redshift of $R_{\rm
  i}^4$, which assuming $R_{\rm i} \approx 10^4$ gives a redshift of
$10^{16}$.  Since {\sc ligo} operates in the KHz region, which
corresponds to energies of the order of $10^7$ GeV, only when the
second burst of inflation occurs at an energy scale larger than this
can we ensure that the {\sc ligo} region will not be affected.
However, in the current scenario, the second epoch of inflation will
be useful only when it occurs at energy scales lower than this.  As
for {\sc lisa}, depending on the scale at which the second epoch of
inflation occurs the situation may not worsen, because {\sc lisa}
scales may remain superhorizon for most or all of the thermal
inflation epoch.  Nevertheless the direct detection of the
gravitational wave background by {\sc lisa} was already impossible in
conventional inflationary models, because the predicted signal is
too low even before thermal inflation does further damage.

\begin{figure}[t]
  \centering \leavevmode\epsfysize=7cm \leavevmode\epsfxsize=7.5cm
  \epsfbox{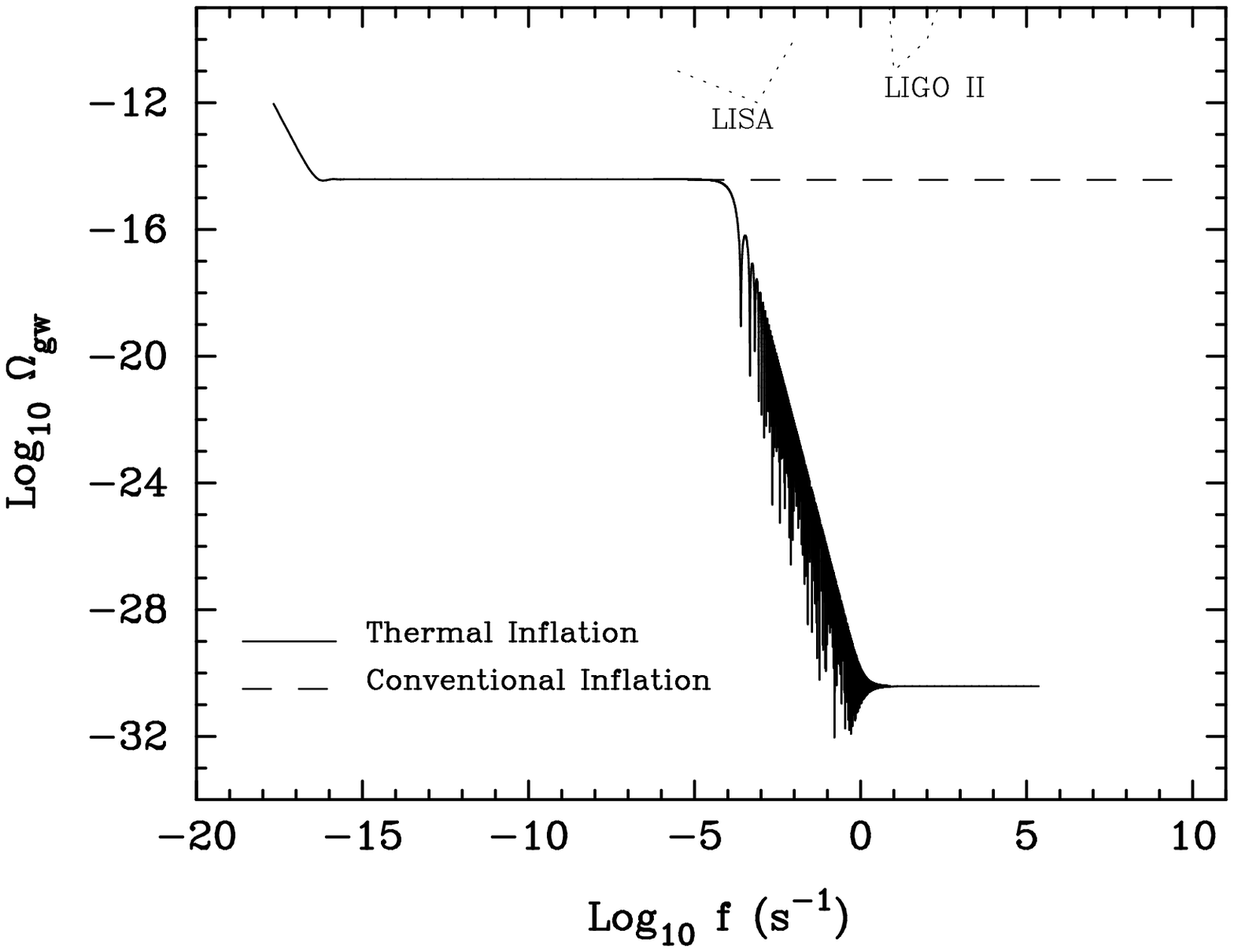} \hspace*{1cm} \leavevmode\epsfysize=7cm
  \leavevmode\epsfxsize=7.5cm\epsfbox{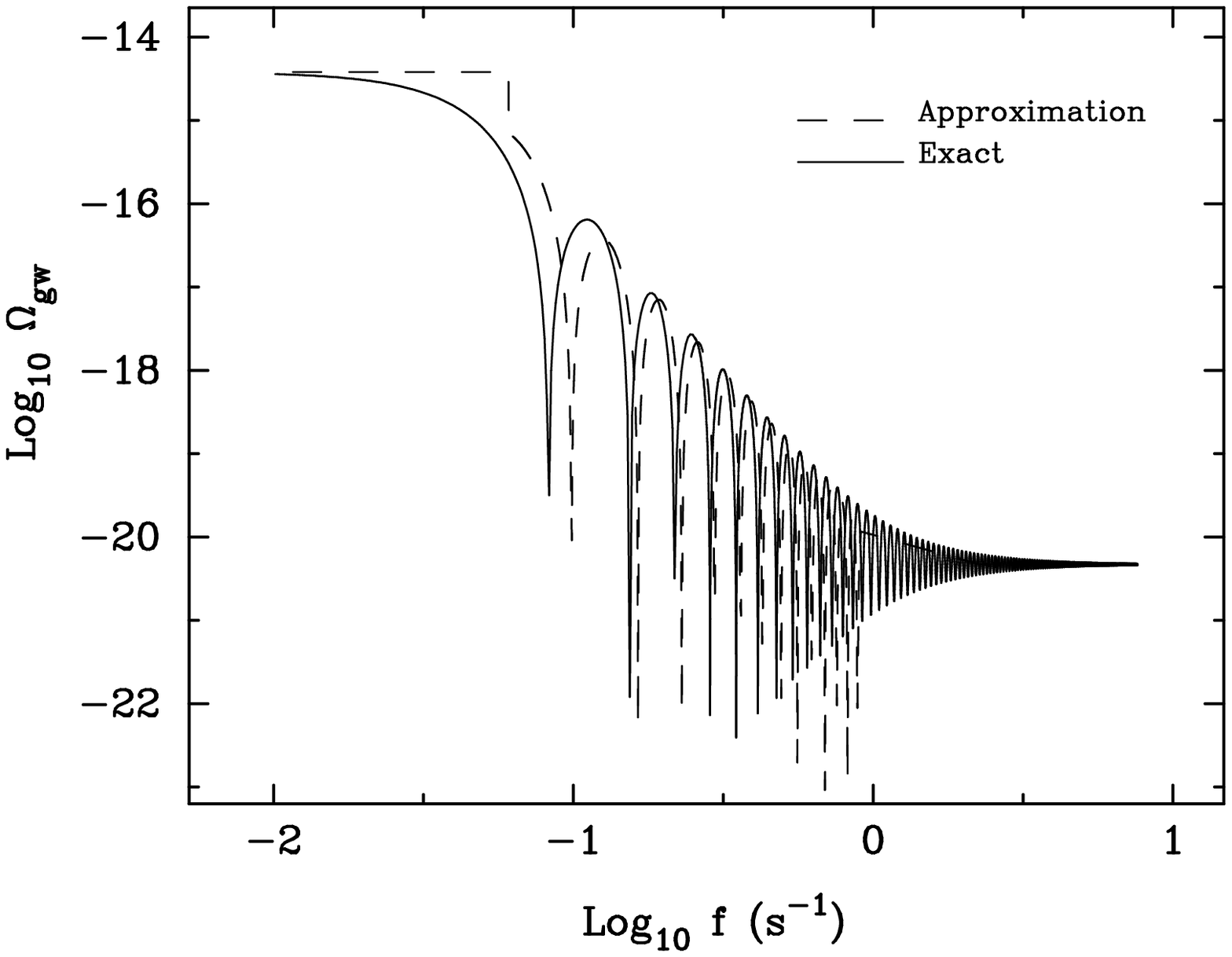}\\
  \caption[fig1]{\label{fig:omegati} The plot on the left shows the 
    exact spectral energy density parameter for the model of thermal
    inflation where we have exponential inflation in both inflationary
    eras. The spectrum for the usual model of exponential inflation
    without an extra epoch of thermal inflation is also shown for
    comparison purposes. The dotted lines marked {\sc lisa} and {\sc
      ligo} II are the expected sensitivities of the gravitational
    wave detectors {\sc lisa} and {\sc ligo} in the advanced
    configuration. Detection was already very problematic in
    conventional models of inflation~\cite{nosee}, and with the
    inclusion of an epoch of thermal inflation the scales relevant for
    {\sc ligo} are redshifted by an additional factor $R_{{\rm i}}^4$
    which makes their detection hopeless. This plot was obtained with
    $\rho_{\rm i}^{1/4}=3 \times 10^{16}$ GeV, $\rho_{{\rm
        i}2}^{1/4}=10^8$ GeV and $R_{\rm i}=10^{4}$. Here, as in all
    the other plots in this paper, we assume $h=0.65$. The plot on the
    right shows a close-up of the oscillations in the central part of
    the spectrum, both for the exact result and the long-wavelength
    approximation given by Eq.~(\ref{eq:omgw_ti}). In order to be able
    to resolve individual oscillations, we used $R_{\rm ti}=30$ rather
    than the realistic value of $10^4$ for this illustration. All the
    other parameters are the same as in the left plot. We checked that
    on increasing the number of $e$-foldings of expansion during
    thermal inflation, the phase difference between the exact result
    and the approximation decreases and for $R_{\rm ti} \approx 100$
    the two curves are almost indistinguishable.}
\end{figure}

The exact result for the spectral energy density parameter when both
the initial and second epoch of inflation (thermal inflation) are
exponential, obtained as described in the Appendix, is plotted in
Fig.~\ref{fig:omegati}. This plot illustrates all the features
discussed so far and also shows that the extra redshift introduced by
thermal inflation makes the direct detection of the stochastic
background almost impossible. In the long-wavelength approximation we
get:
\begin{equation}
\label{eq:omgw_ti}
   \Omega_{{\rm gw}} = \left\{
  \begin{array}{lrcl}
    {\displaystyle \frac{2^{3 p -2}}{3^{p}} \frac{\pi^{1-p} \,
        \Gamma}{p^{2(1+p)}} \left( \frac{ \left( \rho_{\rm c} \,
        \rho_{{\rm r}1}\right)^{p+1}}{\rho_{\rm Pl}^{2 (p+1)}
        \, R_{\rm i}^{4(p-1)} \left( 1 + z_{\rm eq}
        \right)^{p-1}} 
    \right)^{\!\!1/2} 
    \left( \frac{f_{\rm  Pl}}{f} \right)^{\!\!2 p}} \quad ; &  
    H_0 & <f< & {\displaystyle H_0 (1+z_{\rm eq})^{1/2}} \\[5mm]
    {\displaystyle \frac{2^{3 p +1}}{3^{p+1}} \frac{\pi^{2-p} \,
        \Gamma}{p^{2(1+p)}} \left( \frac{\rho_{\rm c}^{p-1}
        \rho_{{\rm r}1}^{p+1}}{\rho_{\rm Pl}^{2 p}
        \, R_{\rm i}^{4(p-1)} \left( 1 + z_{\rm eq}
          \right)^{p+1}}\right)^{\!\!1/2} 
    \left( \frac{f_{\rm  Pl}}{f} \right)^{\!\!2 (p-1)}}; 
  \quad & {\displaystyle H_0 
      (1+z_{\rm eq})^{1/2}} & < f < &
      {\displaystyle \frac{\left( H_{{\rm i}2} H_{0}
          \right)^{1/2}}{R_{\rm i} (1+z_{\rm eq})^{1/4}}}\\[5mm] 
  {\displaystyle \frac{2^{3 (p +1)}}{3^{p+3}} \frac{\Gamma}{p^{2(1+p)}
      \pi^p} \left(\frac{\left( \rho_{\rm c} \,
        \rho_{{\rm r}1}\right)^{(p+1)/2} \rho_{{\rm i}2}}{\rho_{\rm Pl}^{p+2}
        \, R_{\rm i}^{2(p+1)} \left( 1 + z_{\rm eq} \right)^{(p+3)/2}}
    \right)} \times \\[5mm]
    {\displaystyle \hspace{4.9cm} \cos^2\!\left( 2 \pi f \tau_{{\rm
            r}1} \right) 
    \left( \frac{f_{\rm  Pl}}{f} \right)^{\!\!2 (p+1)}}; 
  \quad & {\displaystyle \frac{\left( H_{{\rm i}2} H_{0}
          \right)^{1/2}}{R_{\rm i} (1+z_{\rm
        eq})^{1/4}}} & < f < &
      {\displaystyle \frac{\left( H_{{\rm i}2} H_{0}
          \right)^{1/2}}{\left( 1+z_{\rm eq}\right)^{1/4}}}\\[5mm]
{\displaystyle \frac{2^{3 p +1}}{3^{p+1}} \frac{\pi^{2-p} \,
        \Gamma}{p^{2(1+p)}} \left( \frac{\rho_{\rm c}^{p-1}
        \rho_{{\rm r}1}^{p+1}}{\rho_{\rm Pl}^{2 p} \left( 1+z_{\rm
            eq}\right)^{1+p} R_{\rm i}^{4(p+1)}} \right)^{\!\!1/2} 
    \left( \frac{f_{\rm
          Pl}}{f} \right)^{\!\!2(p-1)}}; 
  \quad & {\displaystyle \frac{\left( H_{{\rm i}2} H_{0}
          \right)^{1/2}}{\left( 1+z_{\rm eq}\right)^{1/4}}} & < f < &
      {\displaystyle \frac{\left( H_{{\rm r}1} H_{0}
          \right)^{1/2}}{R_{\rm i} (1+z_{\rm
        eq})^{1/4}}} \, .
  \end{array}
  \right.
\end{equation}
The cosine term appearing in the third branch of the spectrum deserves
some further explanation. In a full long-wavelength approximation we
would expand the $\cos$ function keeping only the leading order term
which in this case is $1$. However, in this case it can be seen that
$k\tau_{{\rm r}1} > 1$ and the series expansion would not be
appropriate. This is the mathematical reason behind the oscillations
between the two plateaus in Fig.~\ref{fig:omegati}.

The long-wavelength approximation in Eq.~(\ref{eq:omgw_ti}) fits the
exact result very well and it is almost indistinguishable from it
except very close to frequencies associated with a transition and
during the oscillations in the central region of the spectrum. The
right panel in Fig.~\ref{fig:omegati} shows a comparison between the
exact result and the long-wavelength approximation in the oscillating
part of the spectrum. Since the frequency of the oscillations
increases with $\tau_{{\rm r}1} \propto R_{\rm ti}$, in this plot we
used a small value for $R_{\rm ti}$ in order to be able to show the
oscillations in detail. In the particular case shown in the figure,
the oscillations in the approximation have a different phase from
those in the exact result. However, we checked that on increasing
$R_{\rm ti}$ (and therefore the frequency of the oscillations), the
phase difference decreases, until the approximation and the exact
result agree very well for values of $R_{\rm ti} \gtrsim 100$.
 
The factor $R_{\rm i}^{4(p-1)}$ present in the two lower frequency
branches of $\Omega_{\rm gw}$ is due to the change in the
correspondence between comoving and physical scales, as was discussed
above.  We can see that if the initial epoch is exponential inflation
($p=1$) this effect is absent.  This can be understood when we realize
that the effect is a shift in the spectrum towards low frequencies.
When the spectrum is flat, as is the case for exponential inflation,
the net effect for frequencies which were always outside the horizon
is null, while for a tilted spectrum such as the one produced by
power-law inflation this will appear as a further redshift in the
spectrum at these scales.  This means that when the initial spectrum
is generated by power-law inflation and therefore is not flat, the
contribution for the anisotropies in the cosmic microwave background
radiation will be suppressed.  The contribution for the anisotropies
coming from scalar perturbations will also be suppressed, but if the
initial spectra have a different tilt there will be a net effect in
the tensor to scalar ratio.

One can view the effect of thermal inflation as being a `transfer
function' which processes the initial spectrum into a final spectrum.
Its rather complicated form is given by the ratio of the final
Bogolubov coefficient $\beta$ to the initial $\beta$ at the beginning
of the radiation epoch. In the long-wavelength approximation, the
transfer function is independent of the initial spectrum which is
being processed; for example, we expect the suppression to take
exactly the same form in the pre big bang scenario, as we shortly
illustrate.  In the end the transfer function will lead to the same
result as the full long-wavelength approximation (except possibly for
a factor of $O(1)$).  The advantage of the transfer function formalism
is that regardless of how the initial spectrum was produced (in this
subsection conventional inflation, in the next the pre big bang model)
for a given evolution of the Universe after the initial spectrum was
produced, the transfer function is always the same.

Let us now illustrate the use of the transfer function as discussed in
the appendix. Neglecting constants of $O(1)$ the initial spectrum is
in this case given, in terms of the number of particles $N_k$ in mode
$k$ as defined in the appendix, as
\begin{equation}
  \label{eq:init}
  N_k = \frac{1}{(k \tau_{\rm pl1})^{2(p+1)}} \, ,
\end{equation}
where $\tau_{\rm pl1}=2 (\tau_{{\rm r}2} - \tau_{{\rm i}2}) +
\tau_{{\rm r}1}$ is nothing but the argument of the Hankel functions
appearing in the mode functions. From Eqs.~(\ref{eq:transfer}) and
(\ref{eq:transfer_rev}) we easily obtain
\begin{equation}
  \label{eq:tf_ti}
  G\left( k \right) = \left\{
    \begin{array}{lrcl}
      {\displaystyle \tau_{{\rm i}2}^{-2} 
          \tau_{{\rm r}2}^{2} \tau_{\rm eq}^{-2}k^{-2}} & H_0 & < k < & a_{\rm
          eq} H_{\rm eq}\\[1mm]
        {\displaystyle \tau_{{\rm i}2}^{-2} 
          \tau_{{\rm r}2}^{2}} & a_{\rm eq} 
        H_{\rm eq} &< k < & a_{{\rm i}2} H_{{\rm i}2}\\[1mm]
      {\displaystyle \tau_{{\rm r}2}^{-4} k^{-4}} 
      & a_{{\rm i}2} H_{{\rm i}2} & < k < & a_{{\rm
          r}2} H_{{\rm r}2}\\[1mm]
      {\displaystyle 1} & a_{{\rm r}2} H_{{\rm r}2} &< k < & a_{{\rm
          r}1} H_{{\rm r}1}
    \end{array}
  \right.
\end{equation}
After all the factors involving the conformal time at the different
transitions are written in terms of the parameters of our model with
help of Eqs.~(\ref{eq:eq}) to~(\ref{eq:r1}), we obtain the same result
as in Eq.~(\ref{eq:omgw_ti}), except for numerical factors.

\subsection{The pre big bang cosmology}

\label{sec:pbb}

The pre big bang (PBB) is a recently-proposed cosmological scenario
\cite{pbb1,pbb2-frame-equiv} inspired by scale factor duality, one of
the basic ideas of string theory\cite{sfd}. We will briefly summarize
here some of the main features of the PBB scenario. For a more
comprehensive review, covering both the motivation and the
phenomenology, see Ref.~\cite{pbbrev}.

In the PBB scenario the Universe started its evolution from the most
simple initial state conceivable, the perturbative vacuum of string
theory.  In such a state physics can be described by the tree-level
low-energy effective action of string theory given by
\begin{equation}
  \label{eq:steffac_sf}
  S_{\rm eff} = \frac{1}{2 \lambda_{\rm s}^2}\int \sqrt{-g} e^{-\phi} \left[
    {\cal R} + \partial_\mu\phi\partial^\mu\phi - V(\phi) \right] d^4 x
\end{equation}
The details of the potential $V(\phi)$ are still largely unknown,
except that it goes to zero as a double exponential when
$\phi\rightarrow - \infty$ and it must have a non-trivial minimum at
the present.

During an initial stage, the evolution of the Universe is driven by
the kinetic energy of the dilaton field. At this stage both the
curvature scale and the dilaton increase as $t \rightarrow 0_-$, $|H|
\sim |\dot{\phi}| \sim t^{-1}$ and there is in this case an exact
solution for both the scale factor and the
dilaton\cite{pbb1,pbb2-frame-equiv,mode-eq}.
\begin{equation}
  \label{eq:sf_dil}
  \begin{array}{rcl}
    a & = & (-t) ^{-1/2} \; \\[2mm]
    \phi & = &  -3 \ln(-t)\, .
  \end{array}
\end{equation}
The solution with increasing $H$ is in fact an inflationary solution
of the so-called super-inflation or pole-inflation type and this is
one of the main peculiarities of this model. During this stage in the
evolution of the Universe all the couplings will be very small since
they are given by $g=\exp(\phi/2)\approx 0$.

As soon as the curvature reaches the string scale $M_{\rm s} \equiv
\lambda_{\rm s}^{-1}$, all the higher-order derivative terms in the
action become important and the effective action
Eq.~(\ref{eq:steffac_sf}), as well as the solution
Eq.~(\ref{eq:sf_dil}), are no longer valid.  Unfortunately during the
stage where the truly stringy regime is reached not much is known.
What we know however is that during the stringy era the curvature
scale must grow only mildly in order to prevent its unbounded growth
with the consequent singularity (the Big Bang) which we are trying to
avoid in this model. Therefore, whatever the exact form of the
solution is, we expect that $H\approx \textrm{const}$ and
$\dot{\phi}\approx \textrm{const}$. The constraint on $H$ will
therefore lead us to some form of conventional inflation during the
stringy regime.

At some stage during the string phase the dilaton eventually
approaches the strong coupling regime $g \approx 1$ and at this point
the dilaton will freeze, leaving us with the correct value of Newton's
constant, and the transition to conventional cosmology starting with a
radiation-dominated phase will occur.

The evolution we have described so far is based on the action
Eq.~(\ref{eq:steffac_sf}) where the dilaton is non-minimally coupled
to curvature, the so-called string frame. By applying a conformal
transformation involving the dilaton we go from the string frame to
the Einstein frame where we recover the usual Einstein--Hilbert action
for gravity. When we transform from the string frame to the Einstein
frame, the super-inflation type solution for the scale factor is
transformed into accelerated contraction in the Einstein frame.  Since
physics is not changed by a conformal transformation, both frames are
equivalent. After the string era when the dilaton freezes, the string
and the Einstein frames coincide.

One of the most remarkable predictions of the PBB model is a spectrum
of gravitational waves increasing with frequency \cite{pbbgw}. The
reason for this lies in the increase in the curvature during the
dilaton era. Due to this characteristic shape, the cosmological
background of gravitational waves produced in this type of model can
in principle be detected for a broad range of parameters by the
various detectors\cite{pbbgwdet}. We can therefore hope that even when
an epoch of thermal inflation is included in this model, the direct
detection of the gravitational wave background may still be possible
for certain ranges of parameters.

Without loss of generality, we will perform the entire computation of
the gravitational wave background in the string frame, since the
perturbation spectrum is frame-independent~\cite{frame-equiv}.

First of all, we must have an explicit form for the scale factor.
During the string phase there is a well-known solution given in
comoving time by Eq.~(\ref{eq:sf_dil}), which in conformal time $\tau$
takes the form $a \propto \tau^{-1/(1+\sqrt{3})}$ and $\phi \propto -
\sqrt{3}\ln \tau$. During the string era, however, all we know is that
the curvature scale must be approximately constant.  We can therefore
model the stringy phase by an epoch of power-law inflation. The scale
factor before the first radiation dominated era takes the form
\begin{equation}
  \label{eq:pbb}
  a(\tau) = \left\{ 
    \begin{array}{lr}
      {\displaystyle a_{\rm s} \left( \frac{\alpha \: a_{\rm s}}{
            a_{\rm s}^{\prime}} \right)^{\alpha} \left(
          \tau_{\rm dil} - \tau \right)^{-\alpha}}\, ; &  
      -\infty < \tau < \tau_{\rm s} \\[4mm]
      {\displaystyle a_{{\rm r}1} \left( \frac{p \: a_{{\rm r}1}}{
            a_{{\rm r}1}^{\prime}} \right)^{p} \left(
          \tau_{\rm st} - \tau \right)^{-p}} \, ; 
      &  \tau_{\rm s} < \tau < \tau_{{\rm r}1} 
    \end{array}
\right.
\end{equation}
where the subscript `s' denotes the beginning of the string era,
$\alpha=1/(1+\sqrt{3})$, $p$ is a free parameter and
\begin{equation}
  \label{eq:taudef}
  \begin{array}{rcl}
    \tau_{\rm dil}& = & {\displaystyle \tau_{\rm s} + \alpha \frac{a_{\rm
        s}}{a^{\prime}_{\rm s}}} \\[4mm]
    \tau_{\rm st}& = & {\displaystyle \tau_{{\rm r}1} + p \frac{a_{{\rm
          r}1}}{a^{\prime}_{{\rm r}1}}} \, .
  \end{array}
\end{equation}

For the dilaton we have
\begin{equation}
  \label{eq:dilaton}
  \phi(\tau) = \left\{ 
    \begin{array}{lr}
      {\displaystyle \phi_{\rm dil} - \sqrt{3} \ln\left(\tau_{\rm dil}
        - \tau \right)}\, ; &  -\infty < \tau < \tau_{\rm s} \\[4mm]
      {\displaystyle \phi_{\rm st} - 2 \beta \ln\left(\tau_{\rm st}
        - \tau \right)} \, ; & \tau_{\rm s} < \tau < \tau_{{\rm r}1} \\[4mm] 
      {\displaystyle \phi_0} \, ; &  \tau > \tau_{{\rm r}1} 
    \end{array}
\right.
\end{equation}
The constants $\phi_{\rm dil}$ and $\phi_{\rm st}$ are determined by
the continuity of the dilaton, while $\phi_0$ is determined by the
present value of $G$. We will not write them here since they will not
be needed in what follows. As to $\beta$, it is determined by the
continuity of $\phi^{\prime}$ at the transition from the dilaton to
the stringy phase and its value is given by
\begin{equation}
  \label{eq:betast}
  \beta=\frac{\sqrt{3}}{2} \frac{p}{\alpha} = \frac{3+\sqrt{3}}{2} \, p\, .
\end{equation}
This parameter will be important in the determination of the order
$\mu$ of the Hankel functions appearing in the mode functions for the
gravitons during the stringy era (see Appendix) which in this case is
given by
\begin{equation}
  \label{eq:beta_mu}
  \mu = \frac{1}{2} \left| 2 p -2 \beta + 1\right|=\frac{p\left( 1 +
      \sqrt{3} \right)-1}{2}
\end{equation}
After the end of the string phase, a radiation era begins and from
then on the evolution of the Universe is given by
Eq.~(\ref{eq:sf_inf}).

In this model we will have one parameter more than in the context of
conventional cosmology: the total expansion during the string phase,
$R_{\rm s}=a_{{\rm r}1}/a_{\rm s}$.

In this case it is not possible to make $\phi^{\prime}$ continuous at
the transition from the string phase to the radiation phase. This is
the reason why we present the Bogolubov coefficients formalism in the
Appendix in a form which does not require the continuity of the first
derivative of $R$, which in this case is given by $R=a e^{-\phi/2}$.

From now on everything follows as before, except that we must use $R$
instead of $a$, and the expressions for the Bogolubov coefficients
have an extra term due to the discontinuity in $\phi^{\prime}$.

Before we proceed, let us just recall that in the absence of thermal
inflation the spectrum for the gravitational waves is given by
\begin{equation}
  \label{eq:pbb_noti}
   \Omega_{{\rm gw}} \approx \left\{
     \begin{array}{lrcl}
       {\displaystyle \Upsilon \left( \frac{8 \pi}{3} \right)^{1/2} 
         \frac{R_{\rm s}^{\delta}}{\left( 1+z_{\rm 
               eq}\right)^{1/4}} \left( \frac{\rho_{\rm r}^{1/2}
             \rho_{\rm Pl}}{\rho_{\rm c}^{3/2}}\right)^{1/2} \left(
           \frac{f}{f_{\rm Pl}}\right)^3} & {\displaystyle 
         H_0 \left( 1 + z_{\rm eq}
         \right)^{1/2}} & < f < &{\displaystyle \frac{\left( H_{\rm
               r} H_0 \right)^{1/2}}{R_{\rm s}^{1/q} \left( 1 + z_{\rm eq}
         \right)^{1/4}}}\\[5mm]
       {\displaystyle \Theta \left(\frac{2^{6\mu-1} \pi^{5-2\mu}}{3^{2\mu+1}
      q^{2(2\mu+1)}}
  \right)^{1/2} \left( \frac{\rho_{\rm
               c}^{2\mu-3}\rho_{\rm r}^{2\mu+1}}{\rho_{\rm
               Pl}^{2(2\mu-1)} \left( 1+z_{\rm eq} \right)^{2\mu+1}} 
         \right)^{1/4} \left( \frac{f_{\rm Pl}}{f}\right)^{2\mu-3}} & 
       {\displaystyle \frac{\left( H_{\rm
               r} H_0 \right)^{1/2}}{R_{\rm s}^{1/q} \left( 1 + z_{\rm eq}
         \right)^{1/4}}} & < f < &{\displaystyle \frac{\left( H_{\rm
               r} H_0 \right)^{1/2}}{\left( 1 + z_{\rm eq}
         \right)^{1/4}}}
     \end{array}
   \right.
\end{equation}
where
\begin{displaymath}
  \Upsilon = \frac{\left( 1 + \sqrt{3}\right) \left[ p \left( 2 + 
        \sqrt{3}\right) - 1\right]^2 \left[ \pi^2 + \left( 2 + 2
        \gamma - \ln\!4 \right)^2\right]}{
    \left[ p \left( 1 + \sqrt{3}\right) - 1\right]^2}
\end{displaymath}
($\gamma\simeq 0.577$ is the Euler constant) and
\begin{displaymath}
  \Theta=\frac{\left[2\,\Gamma\!\left(1-\mu\right)-\left(
      \left(3+\sqrt{3}\right) q -1
    \right)\Gamma\!\left(\mu\right)\right]^2}{\Gamma^2\!\left(1-\mu\right)
  \Gamma^2\!\left(-\mu\right) }\csc^2\!\left( \mu\pi\right)
\end{displaymath}

Since $\Omega_{\rm gw}$ for modes entering the horizon after the
matter--radiation equality will be far too small to have any
observational consequences, we will forget about this part of the
spectrum. The initial spectrum, i.e. the spectrum at the beginning of
the first radiation era is
\begin{equation}
  \label{eq:in_beta_pbb}
  N_{\rm k}^{(0)}= \frac{\tau_{\rm s}^{2\mu}}{\tau_{\rm r}^{2\mu+1} k}
\end{equation}
Applying the transfer function in Eq.~(\ref{eq:tf_ti}) to this
equation\footnote{We could consider the initial spectrum to be that at
  the beginning of the string epoch. However, in this case the
  transfer function would not be the one in Eq.~(\ref{eq:tf_ti}) since
  it does not take into account the evolution of the gravitational
  waves during the stringy epoch.} and writing all the factors of
conformal time in terms of redshifts and energy scales we obtain:
\begin{equation}
  \label{eq:bogapproxpbb}
  \Omega_{{\rm gw}} \approx \left\{
    \begin{array}{lrcl}
      {\displaystyle \frac{\rho_{{\rm r}1}^{1/4}
          \rho_{\rm Pl}^{1/2}}{\rho_{\rm c}^{3/4}} \frac{R_{\rm
            s}^{\delta} R_{\rm ti}}{\left( 1+z_{\rm eq}\right)^{1/4}}
        \left( \frac{f}{f_{\rm Pl}} \right)^3}\, ; & 
      {\displaystyle H_0 (1+z_{\rm eq})^{1/2}}
      & < f < & {\displaystyle \frac{\left( H_{\rm ti} H_{0}
          \right)^{1/2}}{R_{\rm ti} (1+z_{\rm eq})^{1/4}}}\\[4mm]
      {\displaystyle \frac{\rho_{{\rm r}1}^{1/4} \rho_{\rm c}^{1/4}
          \rho_{\rm ti}}{\rho_{\rm Pl}^{3/2}} \frac{R_{\rm
            s}^{\delta}}{R_{\rm ti} \left( 1+z_{\rm eq}
          \right)^{5/4}} \cos^2\!\left( 2\pi  
          f \tau_{r1} \right) \left( \frac{f_{\rm Pl}}{f} 
        \right)}\, ; & {\displaystyle 
        \frac{\left( H_{\rm ti} H_{0} \right)^{1/2}}{R_{\rm ti} 
          (1+z_{\rm eq})^{1/4}}} & < f < & {\displaystyle 
        \frac{\left( H_{\rm ti} H_{0} 
          \right)^{1/2}}{(1+z_{\rm eq})^{1/4}}}\\[4mm]
      {\displaystyle \frac{\rho_{{\rm r}1}^{1/4} \rho_{\rm
            Pl}^{1/2}}{\rho^{3/4}_{\rm c}} \frac{R_{\rm s}^{\delta}}{R_{\rm
            ti} \left( 1 + z_{\rm eq} \right)^{1/4}} \left(
          \frac{f}{f_{\rm Pl}} \right)^3}\, ; & 
      {\displaystyle \frac{\left( H_{\rm ti} H_{0} 
          \right)^{1/2}}{(1+z_{\rm eq})^{1/4}}} & < f < & {\displaystyle 
        \frac{\left( H_{{\rm r}1} H_{0} \right)^{1/2}}{R_{\rm s}^{1/p}
          R_{\rm ti} \left( 1 + z_{\rm eq} \right)^{1/4}}} \\[4mm]
      {\displaystyle \frac{\rho_{{\rm r}1}^{(\mu+1/2)/2} \rho_{\rm
            c}^{(\mu-3/2)/2}}{\rho_{\rm Pl}^{\mu-1/2} R_{\rm ti}^{2
            \mu+1} \left( 1 + z_{\rm eq}\right)^{(\mu+1/2)/2}}  \left(
          \frac{f_{\rm Pl}}{f} \right)^{2 \mu-3}}\, ; &
      {\displaystyle \frac{\left( H_{{\rm r}1} H_{0}
          \right)^{1/2}}{R_{\rm s}^{1/p} R_{\rm ti} 
          \left( 1 + z_{\rm eq} \right)^{1/4}}}
      & < f < & {\displaystyle \frac{\left( H_{{\rm r}1} H_{0}
          \right)^{1/2}}{R_{\rm ti} 
          \left( 1 + z_{\rm eq} \right)^{1/4}}} 
    \end{array}
  \right.
\end{equation} 
where $\delta=1+\sqrt{3}-1/p$. Here we can see the advantage of using
the transfer function approach instead of applying the long-wavelength
approximation to the full expressions for the Bogolubov $\beta$
coefficient: given the initial spectrum, for a given evolution of the
universe the transfer function only needs to be calculated once.

\begin{figure}[t]
  \centering \leavevmode\epsfysize=7cm \leavevmode\epsfxsize=7.5cm
  \epsfbox{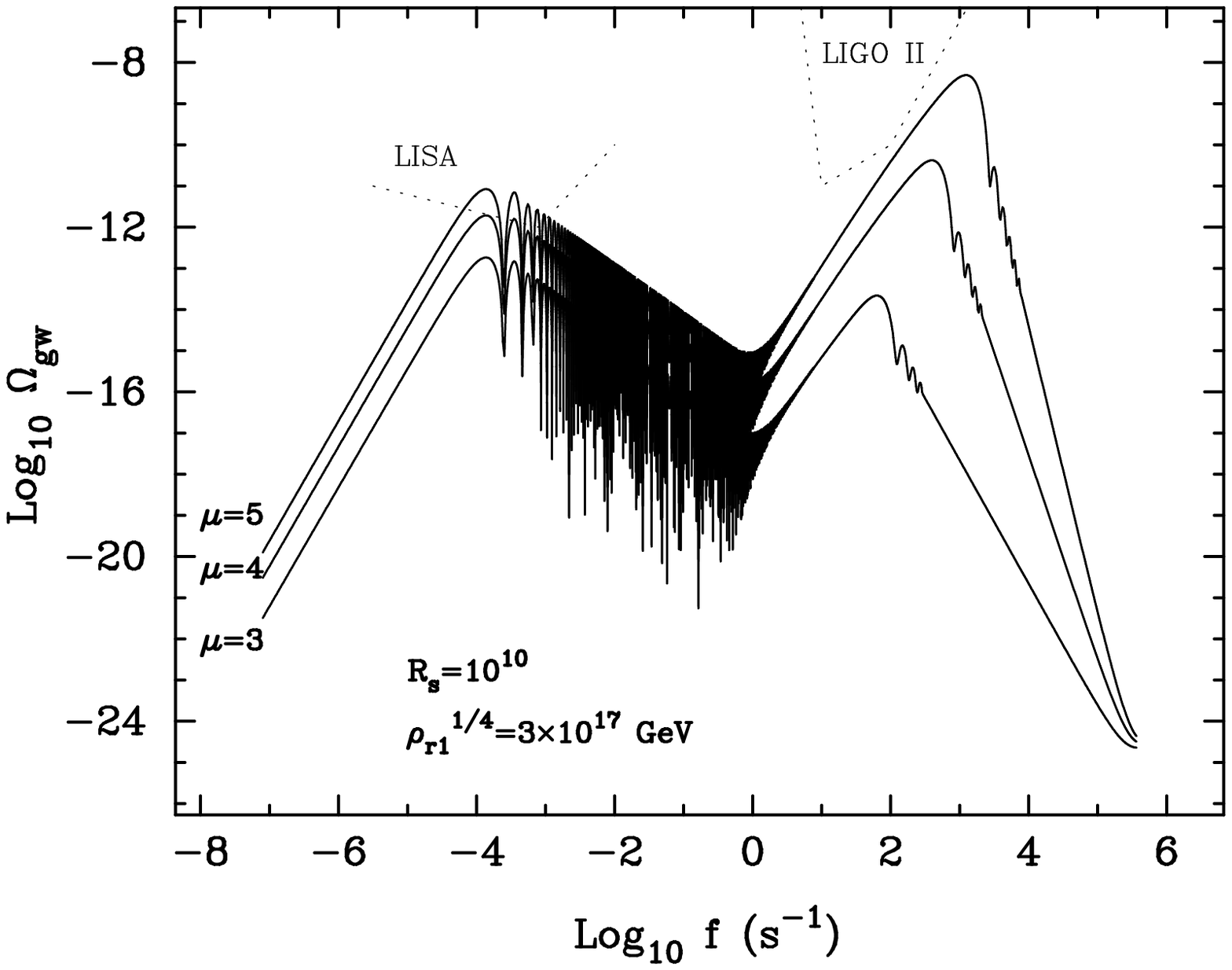} \hspace*{1cm} \leavevmode\epsfysize=7cm
  \leavevmode\epsfxsize=7.5cm \epsfbox{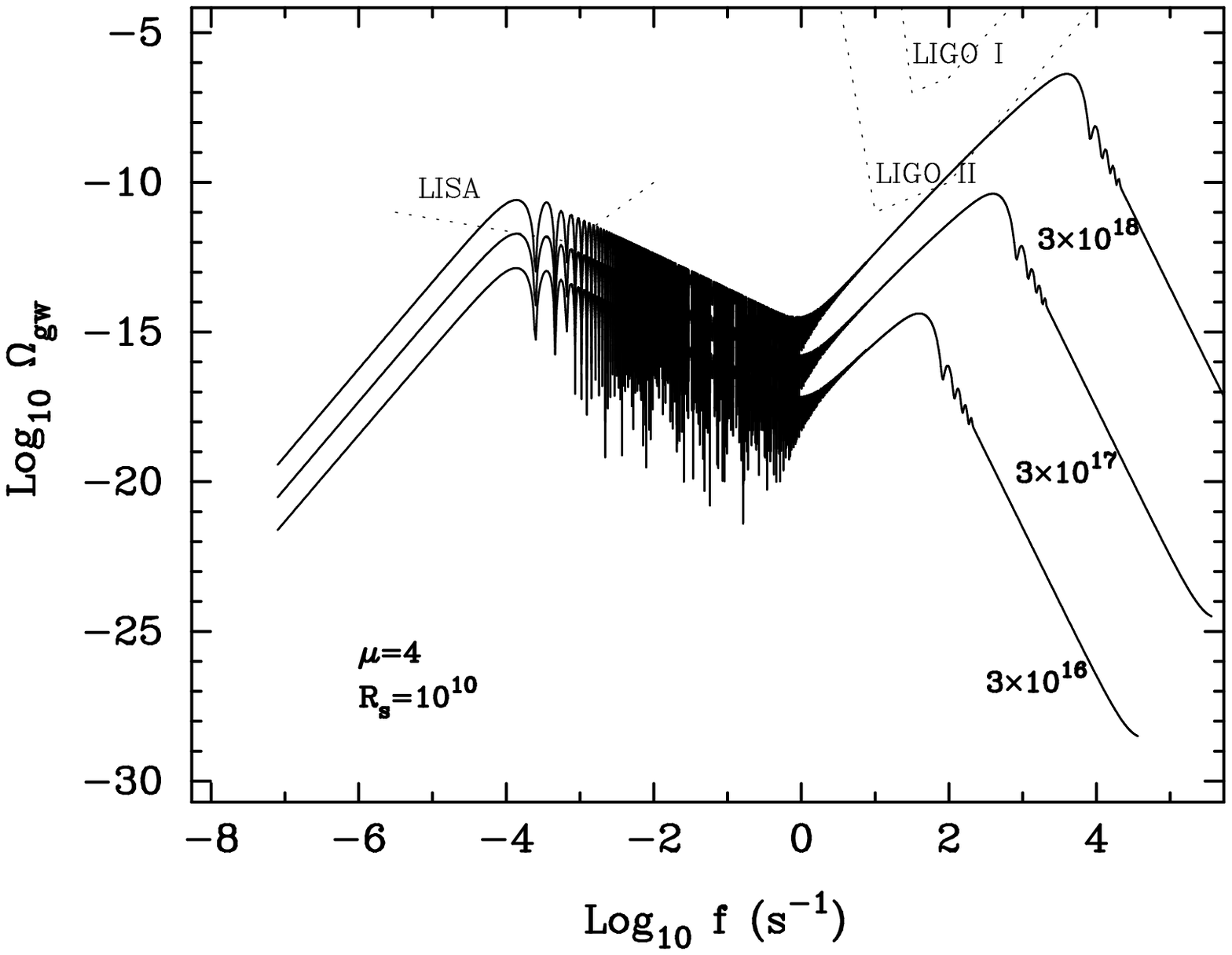} \\
  \caption[fig1]{\label{fig:specpbb} The spectral energy density
    parameter for a model of thermal inflation in the context of PBB
    cosmology. The plot on the left illustrates the dependence of
    $\Omega_{\rm gw}$ on the parameter $\mu$, while the plot on the
    right shows the dependence on the energy scale at the end of the
    stringy phase. Both plots were obtained with $\rho_{\rm
      ti}^{1/4}=10^8$ GeV and $R_{\rm ti}=10^4$. We stress that the
    right-hand peak, near {\sc ligo} scales, violates the conditions
    for a perturbative calculation and is unreliable --- see the text
    for a full discussion.}
\end{figure}

\begin{figure}[t]
  \centering \leavevmode\epsfysize=7cm \leavevmode\epsfxsize=7.5cm
  \epsfbox{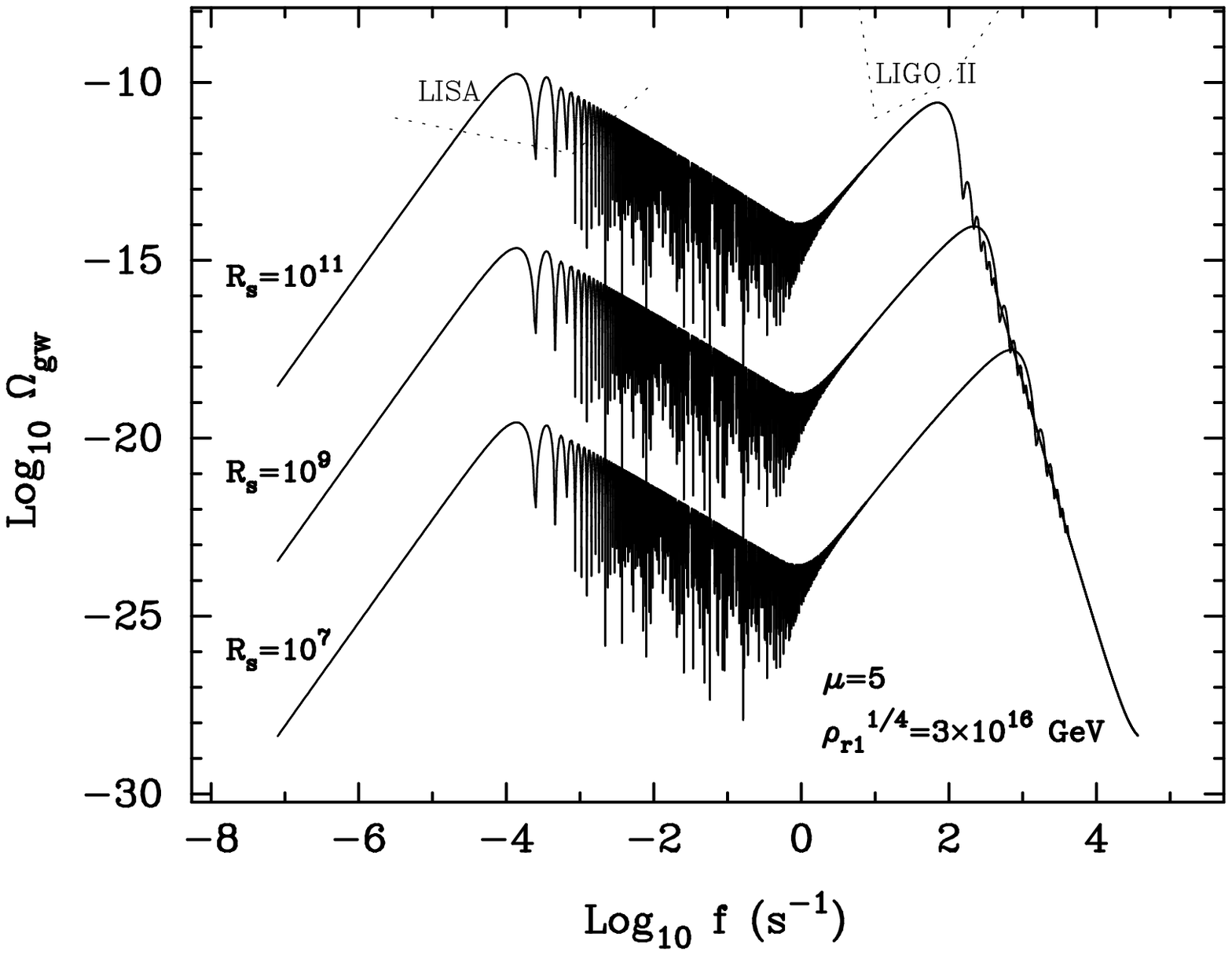} \hspace*{1cm} \leavevmode\epsfysize=7cm
  \leavevmode\epsfxsize=7.5cm \epsfbox{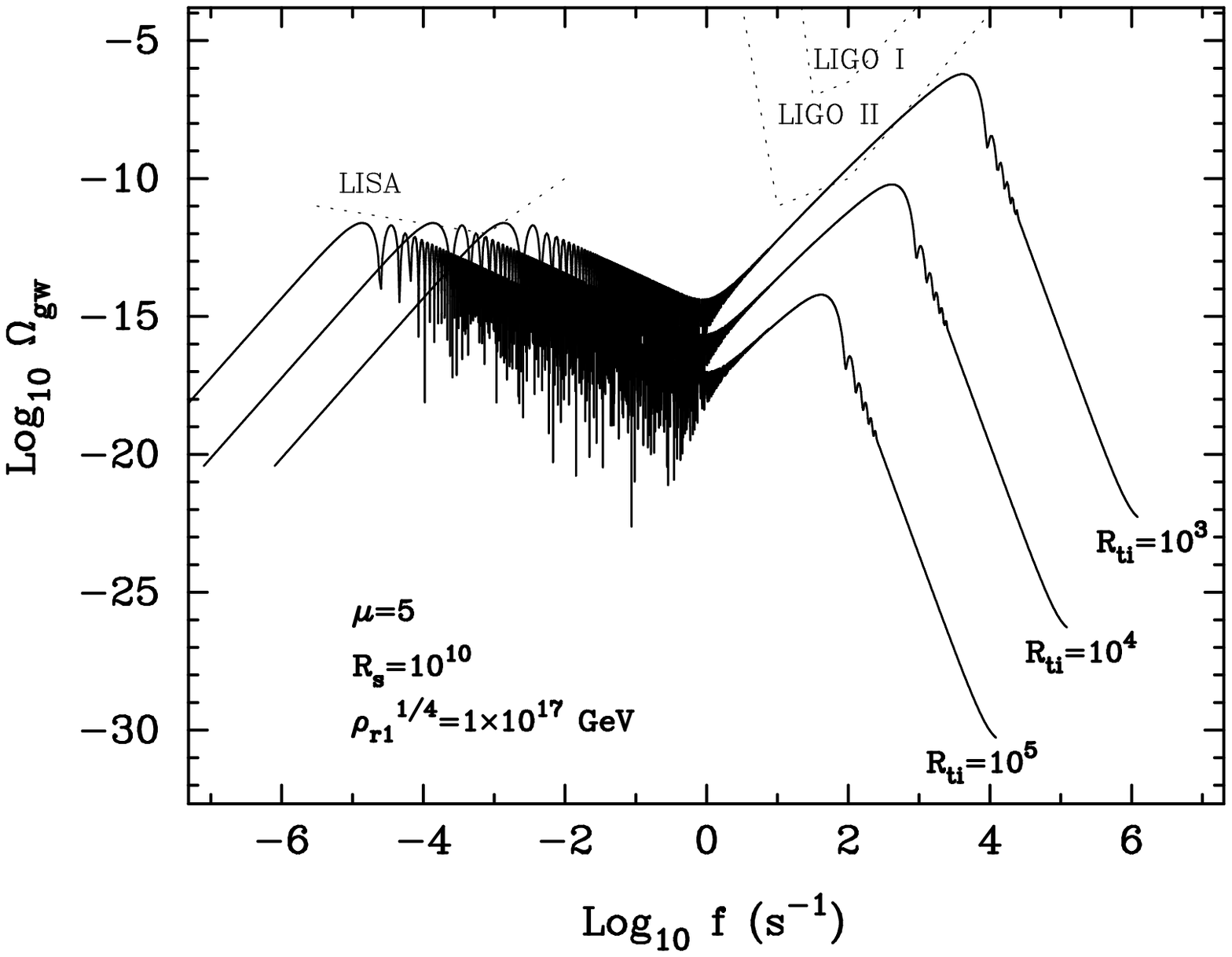} \\
  \caption[fig1]{\label{fig:multspecpbb} The dependence of
    $\Omega_{\rm gw}$ on $R_{\rm s}$ (left) obtained with $\rho_{\rm
      ti}^{1/4}=10^8$ GeV and $R_{\rm ti}=10^4$, and on $R_{\rm ti}$
    for $\rho_{\rm ti}^{1/4}=10^8$ GeV and $R_{\rm s}=10^{10}$.}
\end{figure}

Figs.~\ref{fig:specpbb} and \ref{fig:multspecpbb} show the exact final
spectra and illustrate their dependence on the parameters of the
model. In this model the direct detection of the stochastic background
is still possible in spite of thermal inflation.  As a matter of fact,
because of the energy scale at which the second burst of inflation
starts, the low-frequency peak falls exactly in the frequency range
accessible to {\sc lisa}. This is a happy coincidence and has not
involved any fine tuning in the parameters of the model.

The spectra show a characteristic shape with two peaks, one for
frequencies around $10^2$--$10^4$ Hz and the other in the region
$10^{-2}$--$10^{-4}$ Hz, the exact position of the peaks depending on
the parameters of the model. The high-frequency peak was already
discussed in Ref.~\cite{gall} and it is due to the epoch of power-law
inflation introduced to model the stringy phase. Since the part of the
spectrum produced during the dilaton phase increases with frequency
and the part of the spectrum produced in the stringy phase, which
corresponds to the largest frequencies produced in this model,
decreases with frequency, there must be a peak at the point where the
two curves meet. Although the epoch of power-law inflation introduced
to model the stringy phase is not to be taken very seriously, the fact
is that during the string phase the curvature scale must be
approximately constant, and therefore the part of the gravitational
wave spectrum corresponding to waves produced during this period is
not expected to grow with frequency. We can thus argue that this peak
will most certainly be one of the key signatures in the gravitational
wave spectrum produced by a PBB universe. As to the peak for
frequencies around $10^{-2}$ Hz, it is due to the era of thermal
inflation. The reason for the oscillations is the same as in the
context of conventional cosmology we discussed in the previous
section.

From Fig.~\ref{fig:specpbb}, we see that larger values of $\mu$
(corresponding from Eq.~(\ref{eq:beta_mu}) to larger values of $p$)
correspond to larger values of $\Omega_{\rm gw}$ in the high-frequency
peak. This is because the slope of $\log\Omega_{\rm gw}$ is given by
$2 \mu-3$.

Increasing the energy scale at the end of the string phase shifts the
entire spectrum upwards (Fig.~\ref{fig:specpbb}, right panel).

As to the dependence of $\Omega_{\rm gw}$ on $R_{\rm s}$, we see from
Fig.~\ref{fig:multspecpbb} and the last line of
Eq.~(\ref{eq:bogapproxpbb}) that an increase in $R_{\rm s}$
corresponds to a decrease in the frequency associated with the
dilaton--string transition and an associated increase in the
high-frequency branch of the spectrum which corresponds to the string
era. Associated with this effect there is also the fact that for
$R_{\rm s} > 10^{10}$ the amplitude of the low-frequency peak is
larger than the high-frequency peak, since the frequency associated
with the transition from thermal inflation to radiation is fixed and
the branch associated with the dilaton era ($\propto f^3$) is
therefore shorter.

Finally, the dependence of $\Omega_{\rm gw}$ on $R_{\rm ti}$ is
somewhat unexpected. When all the other parameters are fixed, larger
values of $R_{\rm ti}$ correspond to larger $\Omega_{\rm gw}$ in the
low-frequency branch of the spectrum, contrary to what happens in all
the other branches and what might have been expected, since the epoch
of thermal inflation further redshifts the energy density. Once again
the explanation lies in the behaviour of the frequency associated with
the beginning of the low-frequency branch of the spectrum. Keeping all
the other parameters fixed, this frequency decreases with $1/R_{\rm
  ti}$ therefore leading to a larger increase in the oscillating
branch of the spectrum.

Although the high-frequency peak also falls in the frequency range
accessible to {\sc ligo}, there is a technical subtlety
involved in the derivation of these results which may invalidate our
conclusions regarding the high-frequency peak of the spectrum, which is that
this peak falls outside the domain of application of perturbative
theory as we now demonstrate. Our approach to the calculation of
graviton production is perturbative and therefore we must ensure that
the results obtained satisfy the domain of applicability of the
perturbative calculation, which in our case means that we must have
$h_{ij} \ll 1$. To check the validity of this condition in this
particular case we need to convert our results for the spectral energy
density parameter to the dimensionless amplitude $h$ of the
perturbations.

We can obtain a rough estimate for $h$ by noting that there is an
equivalence between each of the two polarizations of the gravitational
waves, $+$ and $\times$, and a scalar field $\varphi$:
\begin{equation}
\label{eq:grav_equiv}
        h_{+,\times}= \sqrt{16 \pi G} \ \varphi_{+,\times} \ ,
\end{equation} 
from which it follows that
\begin{equation}
\label{eq:h_energy}
        \rho_{\rm gw} \approx \frac{1}{16\pi
        G}\left( \frac{{h^{\prime}}^2}{a^2} + \frac{(\nabla
        h)^2}{a^2}\right) \sim  m_{\rm Pl}^2 \frac{h^2}{\lambda^2} \,,
\end{equation} 
where $\lambda$ is a characteristic wavelength. From Eq.~(\ref{eq:h_energy}) we 
can now obtain the relation we require
\begin{equation}
\label{eq:h_omega}
        h \sim \frac{\sqrt{\rho_{\rm c} \Omega_{\rm gw}}}{m_{\rm
        Pl}} \frac{1}{f} \,.
\end{equation} 
Given the values of $\Omega_{\rm gw}$ and $\rho_{\rm c}$ at a given
time we can estimate the corresponding values of the amplitude $h$. In
the case under study, the problem lies in the high-frequency branch of
the spectrum produced during the transition from the stringy era to
radiation and therefore we must calculate $\Omega_{\rm gw}$ at the
beginning of the radiation era. We can obtain $\Omega_{\rm gw}$ from
the last line of Eq. (\ref{eq:bogapproxpbb}), by taking also into
account the redshift suffered by the frequency $f$ during the
expansion of the Universe:
\begin{equation}
\label{eq:omega_rad}
        \Omega_{\rm gw}^{\rm rad} = R_{\rm ti}^4 (1 + z_{\rm eq})
        \Omega^0_{\rm gw} \,,
\end{equation} 
where $\Omega_{\rm gw}^{\rm rad}$ and $\Omega_{\rm gw}^0$ are the
spectral energy densities of the gravitational wave background at the
beginning of the radiation era and at the present. Using $R_{\rm
  ti}=10^4$, if we have $\Omega_{\rm gw}^0=10^{-11}$, the threshold
for the detection of the high-frequency peak by {\sc ligo}, then
$\Omega_{\rm gw}^{\rm rad}\approx 10^{10}$. This is certainly a very
large value and were we still in the regime where the perturbative
calculation is valid, then this large value for $\Omega_{\rm gw}^{\rm
  rad}$ would represent an inconsistency in our model since we are
assuming $\Omega_{\rm tot}=1$. Using $\rho^{1/4}_{\rm c} \sim 10^{16}$
and noting that frequency is redshifted during the expansion of the
Universe since the end of the stringy epoch by a factor of $10^{31}$,
we obtain from Eq. (\ref{eq:h_omega}) $h_{\rm rad} \approx 10^{9}$ for
frequencies of the order of kHz. This means that our results for the
range of frequencies where we see the first peak are not consistent
with the perturbative approach and therefore they must be taken very
cautiously. However, for the low-frequency peak generated due to the
epoch of thermal inflation no such problem exists, and the
perturbative approach employed here is justified.

This conclusion is completely general; any stochastic background
generated before thermal inflation would have to be non-linearly large
if it were to survive thermal inflation with an amplitude detectable
by {\sc ligo}.

Notice that in this particular model, the requirement that $h\ll 1$ at
the beginning of the radiation era is much stronger than the
constraint $\Omega_{\rm gw} < 10^{-5}$ imposed by
nucleosynthesis\cite{marctes,nsbound}. The nucleosynthesis constraint
is roughly equivalent to requiring $h \lesssim 1$ at the end of
inflation for $f\sim 10^{10}$ Hz which is the high-frequency cutoff.

As a final remark, we should notice that simultaneous detection of the
stochastic gravitational wave background by both {\sc lisa} and {\sc
  ligo} in the advanced phase is not ruled out. As a matter of fact
the frequencies probed by both detectors are appropriate for the
detection of both peaks in the spectrum.

\section{Early matter domination}
\label{sec:earlymat}

Another interesting possible modification to the standard cosmology is
the occurrence of an early epoch of matter domination short after
inflation. A brief early epoch of matter domination may happen during
the QCD phase transition \cite{Schwarz}, but we will concentrate on a
more prolonged epoch, such as might be caused by temporary domination
of the Universe by moduli fields \cite{early-matter}. The scale factor
after the beginning of the first radiation epoch is given by
\begin{equation}
  \label{eq:sf_em}
  a(\tau) = \left\{ 
    \begin{array}{lr}
      {\displaystyle a_2 \tau}\, ; &  
      \tau_{{\rm r}1} < \tau < \tau_{{\rm m}1} \\[4mm]
      {\displaystyle a_3 \left( \tau_3 + \tau \right)^2} \, ; &   
      \tau_{{\rm m}1} < \tau < \tau_{{\rm r}2} \\[4mm]
      {\displaystyle a_4 \left( \tau_4 +\tau \right)} \, ; 
      &  \tau_{{\rm r}2} < \tau < \tau_{{\rm eq}} \\[4mm]
      {\displaystyle a_5 \left(\tau_5 + \tau \right)^2} \, ; &  
      \tau > \tau_{{\rm eq}}
    \end{array}
\right.
\end{equation}
where once again the constants are chosen in order to ensure the
continuity of $a$ and $a^{\prime}$. Notice that in this case the
parameterization of the scale factor is slightly different from the
one in the model with thermal inflation (Eq.~({\ref{eq:sf_inf}})).
Parametrizing the epoch of early matter domination by the energy scale
at which it starts, $H_{{\rm m}1}$, and the total expansion of the
Universe during this era $R_{\rm m}\equiv a_{{\rm r}2}/a_{{\rm m}1}$,
we can write all the parameters of this model in terms of $H_{\rm i}$,
$H_{{\rm m}1}$, $R_{\rm m}$, $H_{0}$ and $z_{\rm eq}$.  The value of
the Hubble parameter at the time of the matter--radiation equality is
in this case also given by Eq. (\ref{eq:eq}) while for $\tau_{\rm eq}$
we have
 \begin{equation}
   \label{eq:teq_rmr}
   \tau_{{\rm eq}} \approx \left( 1
       + z_{{\rm eq}} \right)^{-1/2} H_0^{-1} \, .
\end{equation}
Notice that although the values of $\tau_{\rm eq}$ given by Eq.
(\ref{eq:eq}) for the model of thermal inflation, and by Eq.
(\ref{eq:teq_rmr}) for the model with early matter domination, are the
same, the first equation is exact while the second is an approximation
valid for $z_{{\rm r}2} \gg z_{\rm eq}$, which is always the case in
physically-realistic models. The difference between these two cases
lies in the different parameterization for the scale factor after the
beginning of the second radiation era. For the values of time at the
end of inflation and the beginning and end of the early matter era we
have
\begin{equation}
  \label{eq:t1rmr}
  \tau_{{\rm r}1} = \frac{\left( 1+ z_{\rm eq}\right)^{1/4} R_{\rm
      m}^{1/4}}{\left( H_{\rm i} H_{0}\right)^{1/2}} \ , \quad 
  \tau_{\rm m1} = \frac{\left( 1+ z_{\rm eq}\right)^{1/4} R_{\rm
      m}^{1/4}}{\left( H_{{\rm m}1} H_{0}\right)^{1/2}} \ , \quad
  \tau_{{\rm r}2} \approx \frac{2 \left( 1+ z_{\rm eq}\right)^{1/4} R_{\rm
      m}^{3/4}}{\left( H_{{\rm m}1} H_{0}\right)^{1/2}}
\end{equation}
where the approximation in $\tau_{{\rm r}2}$ is valid for $R_{\rm m}
\gg 1$.

In the case of conventional inflation the scale factor in the initial
epoch is given by Eq.~(\ref{eq:ci}) while in the PBB case the scale
factor is initially given by Eq.~(\ref{eq:pbb}).

The modifications introduced in the spectrum by the epoch of early
matter domination are easy to understand in the light of what has been
said before. First, the high-frequency cutoff will be redshifted.
Since the usual matter era already introduces a factor $(1+z_{\rm
  eq})^{1/4}$ in the high-frequency cutoff, we should also expect the
extra redshift to be of the form $R_{\rm m}^{1/4}$:
\begin{equation}
  \label{eq:hfc2}
  f_{{\rm r}1} = \frac{\left( H_{{\rm i}} H_{0} \right)^{1/2}}{R_{{\rm
        i}1}^{1/4} (1+z_{\rm eq})^{1/4}}  \, .
\end{equation}
 
As before, frequencies which are outside the horizon during the early
matter domination era are not affected by it and we expect the
spectrum to be the same for frequencies smaller than
\begin{equation}
  \label{eq:out_hor_m}
  f_{{\rm r}2} = a_{{\rm r}2} H_{{\rm r}2} = \frac{\left( H_{\rm m}
      H_{0} \right)^{1/2}}{R_{\rm m}^{3/4} 
          (1+z_{\rm eq})^{1/4}} 
\end{equation}

For frequencies in the interval $f_{{\rm m}1} < f < f_{{\rm r}1}$ with
$f_{{\rm m}1} = a_{{\rm m}1} H_{{\rm m}1}$, the spectrum will be
redshifted by a factor $R_{\rm m}$, which corresponds to the ratio
between the energy density of the gravitational waves $\rho_{gw} \sim
a^{-4}$ and the energy density of matter $\rho_{m}\sim a^{-3}$. The
slope of this part of the spectrum is not affected since these waves
have frequencies larger than the frequency cutoff $f_{{\rm m}1}$
introduced by the early epoch of matter domination. For frequencies in
the interval $f_{{\rm r}2} < f < f_{{\rm m}1}$ however, the early
epoch of matter domination will introduce an extra factor of $f^{-2}$
in the frequency dependence. This branch of the spectrum will smoothly
join the redshifted part of the spectrum for high frequencies with the
portion of the spectrum which is not affected by the early epoch of
matter domination. Contrary to the previous case, once a given
wavelength enters the horizon it will remain inside for the rest of
its evolution, and therefore in this case there are no oscillations in
the spectrum.

Using the long-wavelength approximation we obtain
\begin{equation}
\label{eq:omgw_m}
   \Omega_{{\rm gw}} = \left\{
  \begin{array}{lrcl}
    {\displaystyle \frac{2^{3 p -2}}{3^{p}} \frac{\pi^{1-p} \,
        \Gamma}{p^{2(1+p)}} \left( \frac{ \left( \rho_{\rm c} \,
        \rho_{{\rm r}1}\right)^{p+1}}{\rho_{\rm Pl}^{2 (p+1)}
        \, \left(R_{\rm m} \left( 1 + z_{\rm eq} \right)\right)^{p-1}}
    \right)^{\!\!1/2} 
    \left( \frac{f_{\rm  Pl}}{f} \right)^{\!\!2 p}} \quad ; &  
    H_0 & <f< & {\displaystyle H_0 (1+z_{\rm eq})^{1/2}} \\[5mm]
    {\displaystyle \frac{2^{3 p +1}}{3^{p+1}} \frac{\pi^{2-p} \,
        \Gamma}{p^{2(1+p)}} \left( \frac{\rho_{\rm c}^{p-1}
        \rho_{{\rm r}1}^{p+1}}{\rho_{\rm Pl}^{2 p}
        \, R_{\rm m}^{p-1} \left( 1 + z_{\rm eq}
          \right)^{p+1}}\right)^{\!\!1/2} 
    \left( \frac{f_{\rm  Pl}}{f} \right)^{\!\!2 (p-1)}}; 
  \quad & {\displaystyle H_0 
      (1+z_{\rm eq})^{1/2}} & < f < &
      {\displaystyle \frac{\left( H_{{\rm m}1} H_{0}
          \right)^{1/2}}{R_{\rm m}^{3/4} (1+z_{\rm
        eq})^{1/4}}}\\[5mm] 
  {\displaystyle \frac{2^{3 p -2}}{3^{p}} \frac{\pi^{1-p} \,
        \Gamma}{p^{2(1+p)}} \left( \frac{\rho_{\rm c}^{p} \,
        \rho_{{\rm r}1}^{p+1} \rho_{{\rm m}1}}{\rho_{\rm Pl}^{2 (p+1)}
        \, \left(R_{\rm m} \left( 1 + z_{\rm eq} \right)\right)^{p+2}}
    \right)^{\!\!1/2} 
    \left( \frac{f_{\rm  Pl}}{f} \right)^{\!\!2 p}}; 
  \quad & {\displaystyle \frac{\left( H_{{\rm m}1} H_{0}
          \right)^{1/2}}{R_{\rm m}^{3/4} (1+z_{\rm
        eq})^{1/4}}} & < f < &
      {\displaystyle \frac{\left( H_{{\rm m}1} H_{0}
          \right)^{1/2}}{\left( R_{\rm m} (1+z_{\rm
        eq})\right)^{1/4}}}\\[5mm]
{\displaystyle \frac{2^{3 p +1}}{3^{p+1}} \frac{\pi^{2-p} \,
        \Gamma}{p^{2(1+p)}} \left( \frac{\rho_{\rm c}^{p-1}
        \rho_{{\rm r}1}^{p+1}}{\rho_{\rm Pl}^{2 p} \, \left( R_{\rm m}
        \left( 1 + z_{\rm eq} \right)\right)^{p+1}}\right)^{\!\!1/2} 
    \left( \frac{f_{\rm
          Pl}}{f} \right)^{\!\!2(p-1)}}; 
  \quad & {\displaystyle \frac{\left( H_{{\rm m}1} H_{0}
          \right)^{1/2}}{\left( R_{\rm m} (1+z_{\rm
        eq})\right)^{1/4}}} & < f < &
      {\displaystyle \frac{\left( H_{{\rm r}1} H_{0}
          \right)^{1/2}}{\left( R_{\rm m} (1+z_{\rm
        eq})\right)^{1/4}}} \, .
  \end{array}
  \right.
\end{equation}
which confirms our expectations about the modifications introduced by
the early epoch of matter domination. We also see an extra factor of
$R_{\rm m}^{p-1}$ in the two low-frequency branches of the spectrum
which is, as before, due to a change in the correspondence between
comoving and physical scales.

\begin{figure}[t]
  \centering \leavevmode\epsfysize=7cm \leavevmode\epsfxsize=7.5cm
  \epsfbox{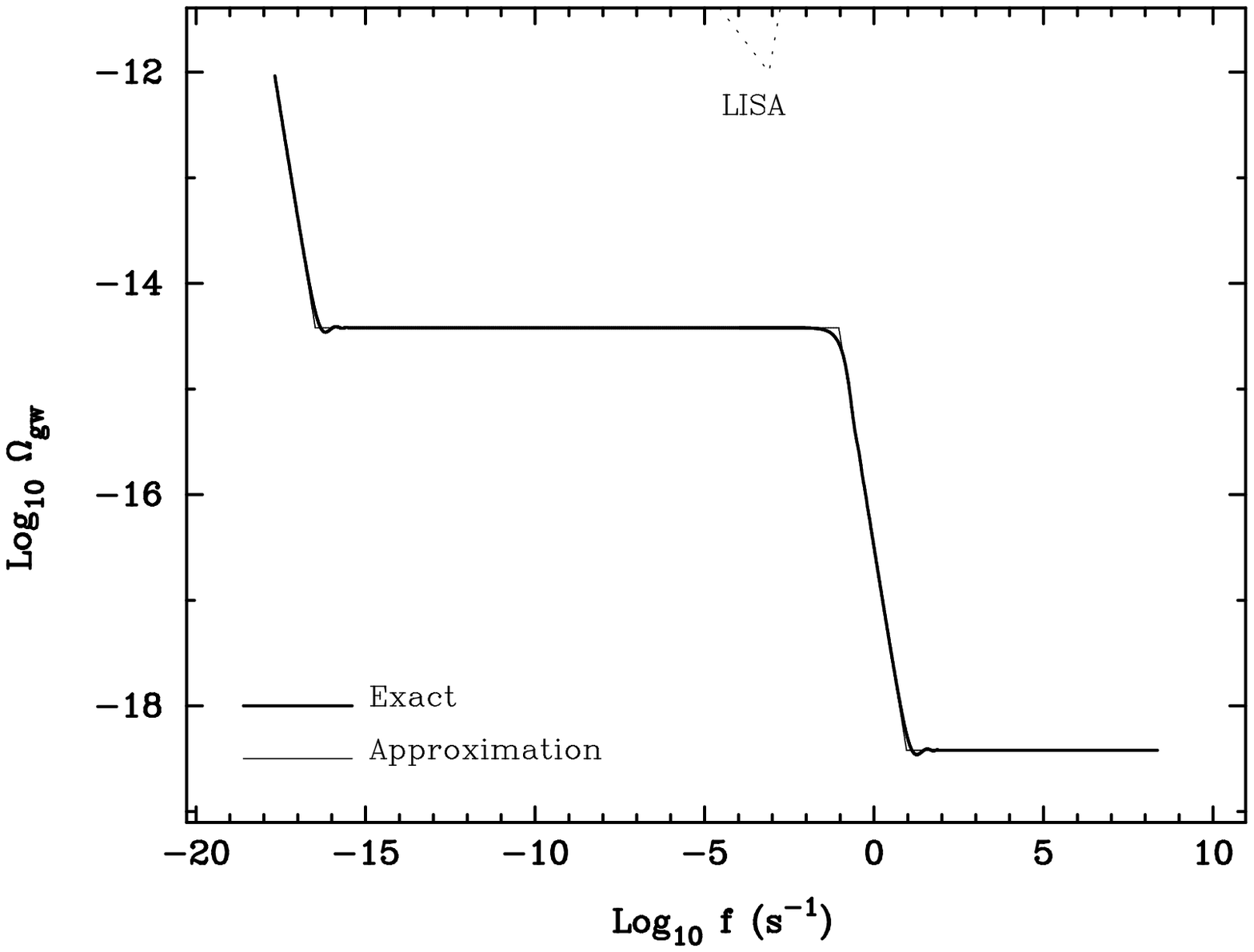}\hspace*{1cm}\leavevmode\epsfysize=7cm
  \leavevmode\epsfxsize=7.5cm
  \epsfbox{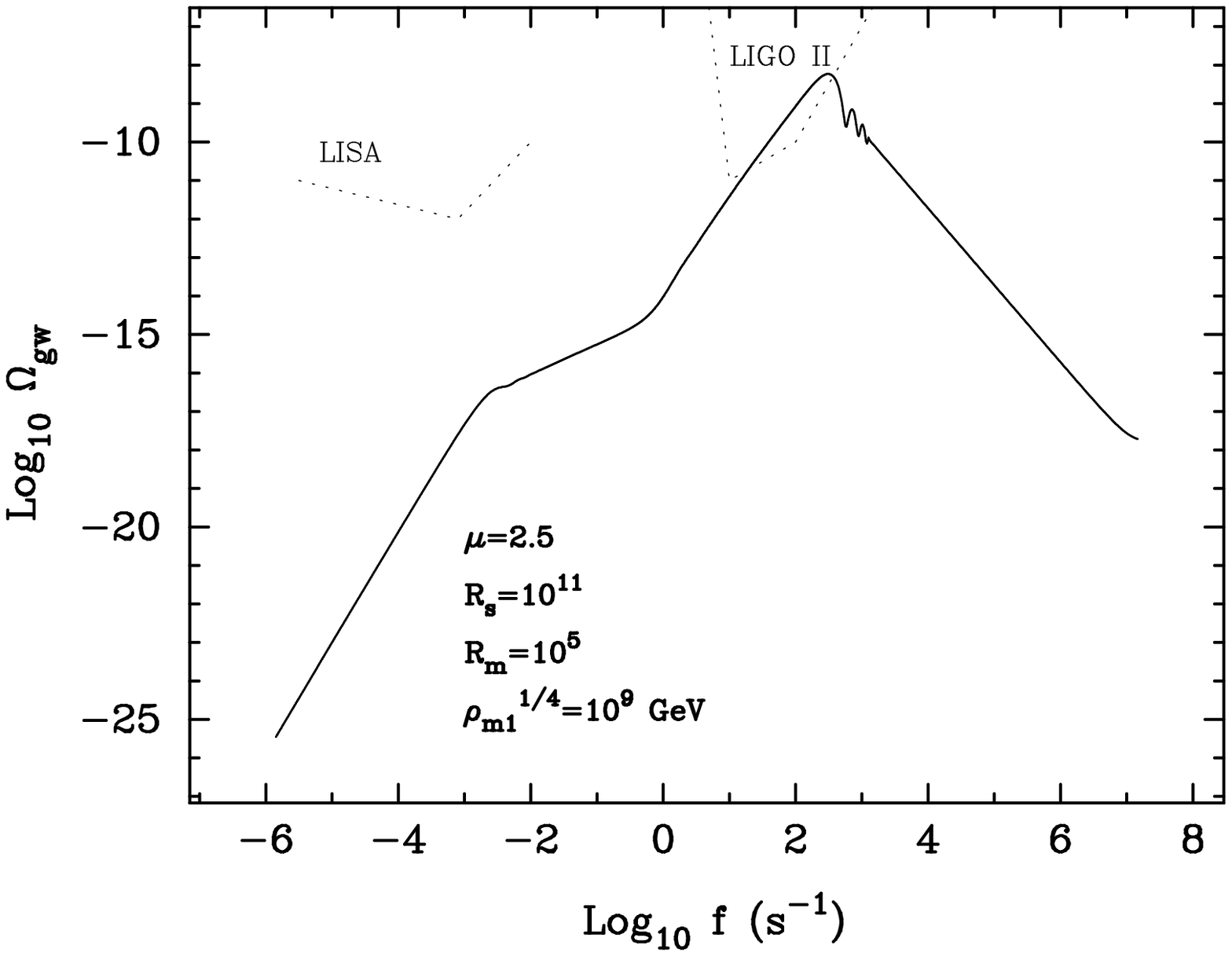} \\
  \caption[fig1]{\label{fig:specrmr} The exact spectral energy
    density parameter for the model with an epoch of early matter
    domination in the context of conventional inflation (left plot)
    shows a suppression for $f > 10$ Hz, but in this case there are no
    oscillations. The suppression caused by the era of early matter
    domination makes the direct detection of the gravitational wave
    background almost impossible. The long-wavelength approximation is
    almost indistinguishable from the exact result except close to the
    transition frequencies. This plot was obtained with $\rho_{\rm
      i}^{1/4}=3 \times 10^{16}$ GeV, $\rho_{{\rm m}1}^{1/4}=10^{10}$
    GeV and $R_{\rm m}=10^{4}$. The right plot is the spectral energy
    density parameter for a PBB model with an epoch of early matter
    domination. Comparing with the case of thermal inflation we see
    the low-frequency peak does not appear in this case. Instead, the
    spectrum grows as $f$ in the intermediate region due to the early
    stage of matter domination. This plot was obtained with
    $\rho_{{\rm r}1}^{1/4}=3 \times 10^{16}$ GeV. For the PBB model,
    the same caveat that the perturbation calculation is unreliable
    applies as for the thermal inflation case.}
\end{figure}

Fig.~\ref{fig:specrmr} shows the exact result for $\Omega_{{\rm gw}}$
in a model with a short period of matter domination where the Universe
started with an epoch of exponential inflation $p=1$. We see that
there is a strong suppression of modes with frequencies larger than
$f_{{\rm mr}}$, the frequency corresponding to the transition from the
epoch of early matter domination to the radiation era. The step
structure is exactly what was found by Schwarz \cite{Schwarz} for the
QCD transition, though much more prominent as our matter-dominated era
is much longer. As in the previous case, the inclusion of an epoch of
early matter domination makes the direct detection of the background
of gravitational waves a very difficult, if not impossible, task.

We have also analyzed a model with early matter domination in the
context of PBB cosmology. Since the details of the model are not
significantly different from those of the models previously analyzed
we will omit them. The final result for the spectrum is shown in the
Fig.~\ref{fig:specrmr} (right plot). In this case the low-frequency
peak is absent and the early epoch of matter domination introduces a
branch which behaves as $f$. Since the slope has the same sign as in
the branch of the spectrum produced by the dilaton era, there is no
peak and detection by {\sc lisa} is not possible. 

As in the previous case, the amplitude of the peak when extrapolated
back in time will give rise to a value of $h \gg 1$ at the beginning
of the radiation phase and once again the results concerning the peak
must be considered very cautiously.

\section{Conclusions}

The unknown cosmological behaviour before nucleosynthesis can
significantly affect the present-day spectrum of cosmological
gravitational waves on scales probed by interferometer experiments. In
this paper we have considered two examples of this effect, one being a
short late burst of inflation and the other a prolonged period of
early matter domination. In each case, the amplitude of gravitational
waves at high frequencies is significantly suppressed compared to the
standard cosmology. In addition, the late inflation model has a very
characteristic signature, in the form of a series of very strong
oscillations in the spectrum.

The influence of these effects on detectability depends on the model
for the initial spectrum. In conventional inflation models the
situation was already extremely pessimistic anyway, so arguably a
further deterioration is of little concern. In the more speculative
pre big bang models of inflation, whose ability to generate observable
gravitational waves has been much remarked upon, the situation is more
complex; while thermal inflation makes it less likely that
gravitational waves might be seen at the {\sc ligo} frequency range,
it opens up a new possibility that they may be seen at the {\sc lisa}
range.  However, the situation is quite model dependent, and there is
the additional worry that on {\sc ligo} scales the usual linear theory
calculations break down, so that predictions on those scales must be
treated with caution. In contrast, our predictions on {\sc lisa}
scales satisfy the linear approximation throughout.

\section*{Acknowledgments}

L.E.M.~was supported by FCT (Portugal) under contract PRAXIS XXI
BPD/14163/97, and A.R.L.~by the Royal Society. We thank Alfredo
Henriques, Jim Lidsey, Gordon Moorhouse and David Wands for useful
discussions, and Bernard Schutz for supplying information on {\sc
  lisa}.

\appendix
\section*{The Bogolubov coefficients formalism}

\label{sec:bogform}

\subsection{Exact results}

\label{sec:bogex}
The mechanism responsible for the generation of the cosmological
background of gravitational waves, parametric amplification of vacuum
fluctuations, was discussed by Grishchuk in the mid
seventies\cite{gris}.  To compute the spectrum of primordial
gravitational waves we use the Bogolubov coefficient
formalism\cite{bogfor}. We assume the transitions between the
different epochs in the evolution of the Universe are instantaneous.
This is in general a very good approximation and it will break down
only for frequencies much larger than $H$ at the instant of the
transition. The accuracy of this approximation has been confirmed in
models where the transitions are not instantaneous\cite{mhm}.

Since in the PBB scenario we will compute graviton production in the
so-called string frame, we summarize this formalism in a form which is
slightly more general than the one most commonly used~\cite{marctes}.
The formalism laid down in this section can be applied to any theory
whose effective action can be written in the form
\begin{equation}
  \label{eq:action}
  S = \frac{1}{{\cal G}}\int \sqrt{-g} \; F\!\left( {\cal R}, \phi
  \right) d^4 x \,,
\end{equation}
where ${\cal G}$ is a constant, $\phi$ is a scalar field and ${\cal
  R}$ is the Ricci scalar. In general relativity ${\cal G}=16 \pi G$
and $F={\cal R}$, while during the dilaton phase in PBB cosmology
${\cal G}=1/2 \lambda_{{\rm s}}^2$, where $\lambda_{{\rm s}}$ is the
string-length parameter, and $F=\exp(-\phi) {\cal R}$ where $\phi$ is
the dilaton. We assume the geometry of the Universe is well described
by a flat Friedmann--Lema\^{\i}tre--Robertson--Walker (FLRW) line
element
\begin{equation}
  \label{eq:flrw}
  ds^2 = dt^2 - a^2(t) d{\bf x}^2 = a^2(\tau) \left( d\tau^2 - d{\bf
      x}^2 \right) \,,
\end{equation}
where $t$ is comoving time and $\tau$ the conformal time obtained by
$d\tau=dt/a$. {}From now on we will always use conformal time.

To first order, the primordial gravitational waves can be represented
as a transverse--traceless perturbation $h_{\mu\nu}$ of the FLRW
metric~\cite{gw}
\begin{equation}
  \label{eq:metric}
  g_{\mu\nu} = g_{\mu\nu}^{(0)} + h_{\mu\nu}  \,,
\end{equation}
where $g_{\mu\nu}^{(0)}$ is the metric of the unperturbed FLRW
background. After quantization, and defining a new quantity
\begin{equation}
  \label{eq:capr}
  R(\tau) \equiv a(\tau) \left( \frac{\partial F}{\partial {\cal R}} 
        \right)^{1/2} \,,
\end{equation}
the $h_{\mu\nu}$ field can be expanded as
\begin{eqnarray}
  \label{eq:gwfield}
  h_{ij}({\bf x},\tau) & = & \sqrt{\cal G} \sum_{\lambda=1}^2 \int
  \frac{d^3k}{(2\pi)^{3/2} R(\tau) \sqrt{2 k}} \chi_{{\bf k},\lambda}({\bf
    x},\tau) \,; \\
  \label{eq:chi}
  \chi_{{\bf k},\lambda}({\bf x},\tau) & = & a_{{\bf k},\lambda}
  \varepsilon_{ij}^{\lambda} e^{i {\bf k}.{\bf x}} \mu_k(\tau) +
  a_{{\bf k},\lambda}^{\dagger}
  {\varepsilon_{ij}^{\lambda}}^{*} e^{- i {\bf k}.{\bf x}} 
  \mu_k^{*}(\tau) \,,
\end{eqnarray}
where $i,j=1,2,3$, the $*$ denotes the complex conjugate, ${\bf k}$ is
the comoving wave vector, $k=|{\bf k}|$ and the sum in
Eq.~(\ref{eq:gwfield}) is over the two polarization states described
by the polarization tensor $\varepsilon_{ij}^{\lambda}$.  The $a_{\bf
  k}^{\dagger}$ and $a_{\bf k}$ (from now on we will drop the
polarization index $\lambda$) are the creation and annihilation
operators for the quanta of the field $h_{\mu\nu}$, the gravitons, and
obey the canonical commutation relations
\begin{equation}
  \label{eq:comutation}
  \left[ a_{\bf k},a_{{\bf k}^{\prime}} \right] = \left[ a_{\bf 
k}^{\dagger}, 
    a_{{\bf k}^{\prime}}^{\dagger} \right] =0 \quad ; \quad \left[ a_{\bf
      k},a_{{\bf k}^{\prime}}^{\dagger} \right] =\delta({\bf
    k}- {\bf k}^{\prime}) \, .
\end{equation}

Inserting the perturbed metric into the Einstein equations, we obtain
the equation which governs the behaviour of the mode functions
$\mu_k$~\cite{marctes,mode-eq}
\begin{equation}
  \label{eq:mueq}
  \mu^{\prime\prime}_k+\left( k^2 - \frac{R^{\prime\prime}}{R}\right)
  \mu_k = 0 \,.
\end{equation}
In most cases of interest, $R \propto \tau^\alpha$ and for such an $R$
the solution of Eq.~(\ref{eq:mueq}) is given in terms of the Hankel
functions $H^{(1)}$ and $H^{(2)}$. Imposing the requirement that in
the limit $\tau \rightarrow - \infty$ we recover the vacuum positive
frequency solution, $\mu_k \rightarrow e^{-i k \tau}$, we can write
$\mu$ as
\begin{equation}
  \label{eq:muhankel}
  \mu_k = \sqrt{\frac{\pi}{2}} \sqrt{k \tau} 
  H^{(2)}_{|\alpha-\frac{1}{2}|}(k \tau) \, .
\end{equation}

During its evolution, the Universe will undergo a series of
transitions between different epochs with different mode functions,
and therefore different vacuum states corresponding to each of these
eras. The relation between the annihilation and creation operators in
the initial state, $a_{\bf k}$ and $a_{\bf k}^{\dagger}$, and in the
final state, $b_{\bf k}$ and $b_{\bf k}^{\dagger}$, is given by a
Bogolubov transformation\cite{bogfor}
\begin{equation}
  \label{eq:bogtransf}
  b_{\bf k} = \alpha_k \, a_{\bf k} + \beta^{*}_k \, a_{\bf - 
        k}^{\dagger} \,,
\end{equation}
where $\alpha_k$ and $\beta_k$ are the Bogolubov coefficients. To
obtain $\alpha_k$ and $\beta_k$ we impose the continuity of the field
$h_{ij}$ and its first derivative at the instant of a transition
between two different eras in the history of the Universe. In general
relativity, where $R\equiv a$, it is always possible to make $R$ and
$R^\prime$ continuous, but since in the context of PBB cosmology it
may not be possible to make $R^\prime$ continuous, as we will see
later, we assume that although $R$ is continuous $R^\prime$ may be
discontinuous at the (instantaneous) transition between two epochs. In
this case, denoting with an overbar quantities referring to the state
after the transition occurred, the Bogolubov coefficients are given
by\cite{marctes}
\begin{eqnarray}
  \label{eq:alphabog}
  \alpha_k & = & \frac{1}{2 i k} \left[ \mu_k \overline{\mu}^{*
      \prime}_k  - \overline{\mu}^{*}_k \mu^{\prime}_k -
    \mu_k \overline{\mu}^{*}_k \frac{\overline{R}^{\prime} -
      R^{\prime}}{R} \right]  \,;\\
  \label{eq:betabog}
  \beta_k & = & \frac{1}{2 i k} \left[ \mu_k^{\prime}
    \overline{\mu}_k - \mu_k \overline{\mu}^{\prime}_k +
    \mu_k \overline{\mu}_k \frac{\overline{R}^{\prime} -
      R^{\prime}}{R} \right] \,,
\end{eqnarray}
where all the quantities are calculated at the time of the transition.
The last term in Eqs.~(\ref{eq:alphabog}) and (\ref{eq:betabog})
appears because we are not imposing the continuity of $R^{\prime}$:
when $R^{\prime}$ is continuous this term is zero and we recover the
usual expression for the Bogolubov coefficients.

By noting that the Wronskian for the solution Eq.~(\ref{eq:muhankel})
of Eq.~(\ref{eq:mueq}) is given by
\begin{equation}
  \label{eq:wrons}
  W = \mu_k \mu^{*\prime}_k - \mu^{*}_k \mu^{\prime}_k = 2 i k \,,
\end{equation}
we obtain an important relation between the Bogolubov coefficients
\begin{equation}
  \label{eq:bogrel}
  |\alpha|^2 - |\beta|^2 = 1 \, .
\end{equation}
The number of particles produced in mode $k$ is given by
\begin{equation}
  \label{eq:nk}
  N_k = |\beta_k|^2 \, ,
\end{equation}
and the spectral energy density parameter is defined by
\begin{equation}
  \label{eq:specenden_def}
  \Omega_{{\rm gw}}\equiv \frac{1}{\rho_{{\rm c}}} \frac{d\rho_{{\rm gw}}}{d 
\ln f} \, ,
\end{equation}
where $f$ is the frequency and $\rho_{{\rm c}}$ and $\rho_{{\rm gw}}$
are, respectively, the critical energy density and the energy density
of gravitational waves, and is given by
\begin{equation}
  \label{eq:specendenpar}
  \Omega_{{\rm gw}} = \frac{\hbar \omega^4}{\pi^2 c^3 \rho_c} N_k \, ,
\end{equation}
where the angular frequency $\omega=2 \pi f=c k/a$.

When we consider two or more transitions in the history of the
Universe, the final Bogolubov coefficients are given by the
composition of the Bogolubov coefficients in each transition,
according to
\begin{eqnarray}
  \label{eq:alphaf}
  \alpha_k & = & \alpha_{1k} \alpha_{2k} + \beta_{1k} \beta_{2k}^{*}  \,;\\
  \label{eq:betaf}  
  \beta_k & = & \beta_{1k} \alpha_{2k}^{*} + \alpha_{1k} \beta_{2k}
  \,.
\end{eqnarray}
From now on we shall drop the index $k$ in $\alpha$ and $\beta$.
Since at any instant modes with wavelengths larger than the horizon
have not yet gone through a complete oscillation, they will not
contribute to the energy density of gravitational waves at that given
time and we are therefore forced to introduce a low-frequency cutoff
for frequencies smaller than
\begin{equation}
  \label{eq:lfcut}
  \omega_{{\rm lf}}(\tau) = 2 \pi H(\tau) \,.
\end{equation}

As we are assuming that all the transitions are instantaneous, a
high-frequency cutoff must also be introduced. In an instantaneous
transition modes with arbitrarily high frequency will be excited.
However, a real transition will take a finite amount of time, and
therefore these high-frequency modes are unphysical, since modes with
frequencies much larger than the rate of expansion of the Universe
will suffer many oscillations during the transition and will not be
affected by it. The value of the high-frequency cutoff is given
by~\cite{al,mar}
\begin{equation}
  \label{eq:hfc}
  \omega_{{\rm hf}}(\tau) = 2 \pi \frac{a_{{\rm tr}} H_{{\rm tr}}}{a(\tau)}
  \,.
\end{equation}
where the subscript `tr' refers to the time when the transition takes
place. This rough argument for the introduction of the high-frequency
cutoff can be formalized in a more rigorous way via Parker's adiabatic
theorem~\cite{par}, which states that modes with frequencies larger
than Eq.~(\ref{eq:hfc}) will be exponentially suppressed. Using a
modification of the formalism presented in this section which allows
for continuous transitions, and introducing a toy model for the
transitions, it is also possible to see that modes with frequencies
larger than the cutoff frequency given in Eq.~(\ref{eq:hfc}) are
indeed exponentially suppressed~\cite{mhm}.

\subsection{Long-wavelength approximation and transfer function}

\label{ap:bogap}

Although the application of the formalism just described is
straightforward, for models involving several transitions the final
result for $\Omega_{\rm gw}$ tends to become very long and not very
instructive, and an approximation is due at this point. The high-frequency
cutoff given by Eq.~(\ref{eq:hfc}) can be written in comoving
coordinates as
\begin{equation}
  \label{eq:k_hfc}
  k_{\rm hf} =  2 \pi a_{\rm tr} H_{\rm tr}\, .
\end{equation}
This corresponds to frequencies which are just entering the horizon at
a given transition. Since $k < k_{\rm hf}$ and $k_{\rm hf}\tau \approx
(a^{\prime}/a) \tau \approx 1$, then $k \tau < 1$ in most cases of
interest and we can replace the Hankel functions appearing in the mode
functions (Eq.~(\ref{eq:muhankel})) by their leading terms. This
approximation is usually very good for most purposes as we can see in
Fig.~\ref{fig:omegati}. Since modes satisfying $k\tau < 1$ have
wavelengths larger than the Hubble radius, this approximation is known
as the long-wavelength approximation.

The long-wavelength approximation can be used to generate a transfer
function for the gravitational wave spectrum. Once we have an initial
spectrum then, for a given subsequent evolution of the Universe, the
same transfer function can always be applied to the initial spectrum,
no matter what its form or origin is, to produce the spectrum at the
present.

Noting that the leading term in the expansion of the Hankel function
$H^{(2)}$ is given by
\begin{equation}
  \label{eq:hankexp}
   H^{(2)}_{\nu}(k \tau) = \frac{i \, 2^{\nu} \csc\!\left(
       \nu\pi\right)}{\Gamma\!\left( 1-\nu\right)}
   \frac{1}{(k\tau)^{\nu}} + \cdots \,,
\end{equation} 
from Eq.~(\ref{eq:muhankel}) we can immediately see that in the long
wavelength approximation the mode functions are given by
\begin{equation}
  \label{eq:muexp}
  \mu \propto \left( k \tau\right)^{-\nu+1/2} \, .
\end{equation}

After a succession of $n$ transitions we should expect the final
$\beta$ to be given by a product of terms of the form
$(k\tau_1)^{\nu_{1}}\cdots (k\tau_{\rm n})^{\nu_{\rm n}}$ with the
possibility that some of the terms may repeat themselves in the
product. After we replace the mode functions by their long-wavelength
approximation, from Eqs.~(\ref{eq:alphabog}), (\ref{eq:betabog}),
(\ref{eq:alphaf}), (\ref{eq:betaf}) we get
\begin{displaymath}
  \beta \approx (k\tau_1)^{-\nu_1} \left(
    \frac{\tau_1}{\tau_2}\right)^{-\nu_2} \cdots \, 
  \left( \frac{\tau_{{\rm n}-1}}{\tau_{\rm n}} \right)^{-\nu_n}
  (k\tau_{\rm n})^{-\nu_{{\rm n}+1}} \, , \quad k_{{\rm n}+1} < k < k_{\rm n}
\end{displaymath}
where we dropped all the numerical factors. The index $n+1$ appears in
the last term since the $n$th transition links the epochs $n$ and
$n+1$. The number of particles for the branch of the spectrum
corresponding to the $n$th transition will therefore be given by
\begin{equation}
  \label{eq:tfapp}
  N_{\rm k} = \left| \beta\right|^2 \approx  \tau_1^{-2(\nu_1+\nu_2)}
  \tau_2^{-2(\nu_3-\nu_2)} \cdots \tau_{{\rm n}-1}^{-2(\nu_{\rm
      n}-\nu_{{\rm n}-1})} \tau_{\rm n}^{-2(\nu_{{\rm n}+1}-\nu_{\rm
      n})} \, k^{-2(\nu_{{\rm n}+1}+\nu_1)}\, ,\quad  
k_{{\rm n}+1} < k < k_{\rm n} \, .
\end{equation}

From Eqs.~(\ref{eq:alphabog}), (\ref{eq:betabog}), (\ref{eq:alphaf}),
(\ref{eq:betaf}) we see that the final Bogolubov coefficients will be
a sum of products of Hankel functions, each term with $2 n$ of these,
in the case of the $n$th transition. We can therefore ask ourselves
why the final dependence on $k$ is just $k^{-2(\nu_{{\rm
      n}+1}+\nu_1)}$ and does not involve any of the exponents
$\nu_2\cdots\nu_{{\rm n}-1}$ which should be contributed by the
previous transitions. The reason for this cancellation can be
understood by noticing that
\begin{displaymath}
  H^{(2)}_{\nu}(x) = J_{\nu}(x)- i \, Y_{\nu}(x) 
\end{displaymath}
where $J$ and $Y$ are the Bessel functions of $1$st and $2$nd kind.
The $Y$ function contributes the leading term $x^{-\nu}$ while to
lower order $J \sim x^{\nu}$. The intermediate powers of $k$ cancel
because $J \sim x^{\nu}$ and in the end we are left with
$\beta\propto k^{-(\nu_1+\nu_{{\rm n}+1})}$.

We can use Eq.~(\ref{eq:tfapp}) to build a transfer function for the
evolution of the gravitational waves. The initial spectrum (the
spectrum after the first transition) takes the form
\begin{equation}
  \label{eq:beta_init}
  N_{\rm k}^{(0)} = \tau_1^{-2(\nu_1+\nu_2)} k^{-2(\nu_1+\nu_2)} \, .
\end{equation}
We are looking for a transfer function, which we shall call $G(k)$,
such that $N_{\rm k} = G(k) N_{\rm k}^{(0)}$. From
Eq.~(\ref{eq:tfapp}) we can easily see that the required $G(k)$ is
given by:
\begin{equation}
  \label{eq:transfer}
  G(k) = \left\{
    \begin{array}{lrcl}
      {\displaystyle \tau_2^{-2(\nu_3-\nu_2)}\cdots\tau_{{\rm
            n}-1}^{-2(\nu_{\rm n}-\nu_{{\rm n}-1})} \tau_{\rm
            n}^{-2(\nu_{{\rm n}+1}-\nu_{\rm n})}k^{-2(\nu_{{\rm 
      n}+1}-\nu_2)}} &  k_{{\rm n}+1} & < k < & k_{\rm n}\\[1mm]
      {\displaystyle
        \tau_2^{-2(\nu_3-\nu_2)}\cdots\tau_{{\rm n}-1}^{-2(\nu_{\rm
      n}-\nu_{{\rm n}-1})} k^{-2(\nu_{\rm
      n}-\nu_2)}} & k_{\rm n} & < k < & k_{{\rm n}-1} \\
       \qquad \cdots & & \cdots & \\
     {\displaystyle \tau_2^{-2(\nu_3-\nu_2)} k^{-2(\nu_3-\nu_2)}} 
      & k_3 & < k < & k_2 \\
      {\displaystyle 1} & k_2 & < k < & k_1 
    \end{array}
  \right.
\end{equation}
Note that $G(k)$ is independent of the
initial spectrum and can therefore, once calculated for a given
cosmology, be applied to a variety of different choices of initial
spectrum. Usually, due to requirements of continuity of the scale
factor, the time dependence of $a$ during the $n$th epoch is given in
terms of $\tau_{\rm n}^{\prime} + \tau$ where $\tau_{\rm n}^{\prime}$
is a constant. As a consequence the argument of the Hankel functions
will be $k(\tau_{\rm n}^{\prime} + \tau)$ and therefore, in
Eq.~(\ref{eq:transfer}) $\tau_{\rm n}$ must be replaced by $\tau_{\rm
  n}^{\prime} + \tau_{\rm n}$.

In our model of thermal inflation there is however a further
complication: the cutoffs associated with the transitions from the
first epoch of radiation to the epoch of thermal inflation, $k_{\rm
rti}$, and from thermal inflation to the final radiation epoch,
$k_{\rm tir}$, satisfy $k_{\rm rti}< k_{\rm tir}$, contrary to the
common situation where later transition will give rise to cutoffs at
smaller frequencies. Assuming this situation involves transitions between 
branches $i$ and $i+1$ and between $i+1$ and $i+2$ ($i>1$), the transfer 
function will
take the form
\begin{equation}
  \label{eq:transfer_rev}
  G(k) = \left\{
    \begin{array}{lrcl}
      \qquad \cdots & & \cdots & \\
      {\displaystyle
        \tau_2^{-2(\nu_3-\nu_2)}\cdots\tau_{{\rm
            i}-1}^{-2(\nu_{\rm i}-\nu_{{\rm i}-1})}
        \tau_{\rm i}^{-2(\nu_{{\rm
      i}+1}-\nu_{\rm i})}\tau_{{\rm
            i}+1}^{-2(\nu_{{\rm i}+2}-\nu_{{\rm i}+1})} k^{-2(\nu_{{\rm
      i}+2}-\nu_2)}} & k_{{\rm i}+2} & < k < & k_{\rm i} \\[1mm]
     {\displaystyle \tau_2^{-2(\nu_3-\nu_2)}\cdots\tau_{{\rm
            i}-1}^{-2(\nu_{\rm i}-\nu_{{\rm i}-1})} \tau_{{\rm
            i}+1}^{-2 (\nu_{{\rm i}+1}+\nu_{{\rm i}+2})}
        k^{-2(\nu_{\rm i}+\nu_{{\rm i}+1}+\nu_{{\rm i}+2}-\nu_2)}} 
      & k_{\rm i} & < k < & k_{{\rm i}+1} \\[1mm]
     {\displaystyle \tau_2^{-2(\nu_3-\nu_2)}\cdots\tau_{{\rm
            i}-1}^{-2(\nu_{\rm i}-\nu_{{\rm i}-1})} k^{-2(\nu_{\rm 
      i}-\nu_2)}} &  k_{{\rm i}+1} & < k < & k_{{\rm i}-1}\\
       \qquad \cdots & & \cdots & 
    \end{array}
  \right.
\end{equation}
Obviously, since here we are taking only the lowest-order terms, with
this approximation we cannot recover the oscillations seen in
Figs.~\ref{fig:omegati}, \ref{fig:specpbb} and
\ref{fig:multspecpbb}, but at least we can get the correct dependence
on frequency.

 
\end{document}